\documentclass[aps,twocolumn,showpacs,preprintnumbers,amsmath,amssymb,superscriptaddress,floatfix,nofootinbib]{revtex4}
\usepackage{lipsum}
\usepackage{graphicx}
\usepackage{epsfig}
\usepackage{epstopdf}
\usepackage{hyperref}
\usepackage{amsmath}
\usepackage{amsfonts}
\usepackage{amssymb}
\usepackage{setspace}
\usepackage{amsfonts,color}
\usepackage{natbib}

\begin{document}

\title{Triangle mechanisms in the build up and decay of the $N^*(1875)$ }
\author{Daris Samart}
\email{daris.sa@rmuti.ac.th}
\affiliation{Department of Applied Physics, Faculty of Sciences and Liberal arts, Rajamangala University of Technology Isan, Nakhon Ratchasima, 30000, Thailand}
\affiliation{Center Of Excellent in High Energy Physics \& Astrophysics, Suranaree University of Technology, Nakhon Ratchasima, 30000, Thailand}
\affiliation{Departamento de
F\'{\i}sica Te\'orica and IFIC, Centro Mixto Universidad de
Valencia-CSIC Institutos de Investigaci\'on de Paterna, Aptdo.
22085, 46071 Valencia, Spain}

\author{Wei-Hong Liang}
\email{liangwh@gxnu.edu.cn}
\affiliation{Department of Physics, Guangxi Normal University,
Guilin 541004, China}

\author{Eulogio Oset}
\email{oset@ific.uv.es}
 \affiliation{Departamento de
F\'{\i}sica Te\'orica and IFIC, Centro Mixto Universidad de
Valencia-CSIC Institutos de Investigaci\'on de Paterna, Aptdo.
22085, 46071 Valencia, Spain}

\date{\today}

\begin{abstract}
We have studied the $N^*(1875) (3/2^-)$ resonance with a multichannel unitary scheme, considering the $\Delta \pi$ and $\Sigma^* K$, with their interaction extracted from   chiral Lagrangians, and then have added two more channels, the $N^*(1535) \pi$ and $N \sigma$, which proceed via triangle diagrams involving the  $\Sigma^* K$ and $\Delta \pi$ respectively in the intermediate states. The triangle diagram in the $N^*(1535) \pi$ case develops a singularity at the same energy as the resonance mass. We determine the couplings of the resonance to the different channels and the partial decay widths. We find a very large decay width to $\Sigma^* K$, and also see that, due to interference with other terms, the $N \sigma$ channel has an important role in the $\pi \pi$ mass distributions at low invariant masses, leading to an apparently large $N \sigma$ decay width. A discussion is done, justifying the convenience of an experimental reanalysis of this resonance to the light of the findings of the paper, using multichannel unitary schemes.
\end{abstract}

\maketitle

\section{Introduction}
The $N^*$(1875) ($3/2^-$) is relatively new in the PDG \cite{pdg}. Quoting from the latest PDG edition, ``Before the 2012 Review, all evidence for a $J^P = 3/2^-$ state with a mass above 1800 MeV was filed under a two star $N$(2080). There is now evidence from Ref.~\cite{anisovich} for two $3/2^-$ states in this region, so we have split the older data (according to mass) between a three-star $N$(1875) and a two-star $N$(2120)." The mass of the PDG is 1820-1920 MeV (1875 MeV PDG estimate) and the width 250 $\pm$ 70 MeV. Quoting directly from Ref.~\cite{anisovich}, the mass is 1880 $\pm$ 20 MeV and the the width 200 $\pm$ 70 MeV. A more recent experiment \cite{sokhoyan} agrees with these values with 1875 $\pm$ 20 MeV for the mass and 200 $\pm$ 25 MeV for the width. The most important decay modes are $N\omega$ (15-25\%), $\Delta(1232)\pi$ (10-35\%), mostly in $s$-wave, and $N\sigma$ ($Nf_0(500)$) (30-60\%).

It is interesting to recall that prior to its acceptance as a new resonance, a peak in the amplitudes was observed around 1875 MeV from the study of the pseudoscalar meson-baryon decuplet interaction in Ref.~\cite{sarkar}. For the case of strangeness $S=0$ and isospin $I=\frac{1}{2}$\,, the coupled channels $\Delta\pi$ and $\Sigma^*K$ were used, and the interaction was obtained from the meson-baryon Lagrangians of Ref. \cite{manohar}. The peak appears at the $\Sigma^*K$ threshold and it was identified as a threshold effect, not a genuine resonance. One should note that the identification of threshold effects with resonances is quite common and one has a good example with the $a_0$(980) which is catalogued as a resonance, but it shows both theoretically \cite{npa} and experimentally \cite{kornicer} as a cusp effect with no clear pole associated to it.

In the present paper we retake the work of Ref. \cite{sarkar} and include triangle mechanisms associated to the main building channels $\Delta\pi$, $\Sigma^*K$, which lead to new channels $N^*(1535)\pi$
and $Nf_0(500)$. The first channel has not been measured yet, but the second channel is, together with $\Delta\pi$ the main decay channel of the resonance. An effective transition potential is constructed from the $\Delta\pi$, $\Sigma^*K$ channels to the $N^*(1535)\pi$ and $Nf_0(500)$, and a four channel problem is then solved with a unitary coupled channel scheme, leading to a resonant peak around 1875 MeV in the amplitudes from where we extract the coupling of the $N^*(1875)$ resonance to the different channels and the partial decay widths to these channels.

Triangle diagrams have for long been part of hadron physics, but of particular interest are those that lead to singularities in the amplitudes, known as triangle singularities. The concept and detailed study was introduced by Landau \cite{landau}, but it is precisely now, after much information on the hadron spectrum and reactions has been accumulated, that many examples of triangle singularities show up \cite{qzhao}. A triangle diagram stems from a particle A decaying into $1+2$, particle 2 decaying to $B+3$ and particles $1+3$ merging into another particle C. In some cases,
when the process can occur at the classical level, a singularity appears in the corresponding Feynman diagram, Coleman-Norton theorem \cite{ncol}, and the field theoretical amplitude becomes infinity if the intermediate particles are stable. In practice, some of these particles have a finite width and the infinity becomes a peak, with important experimental consequences.

An alternative formulation to the standard method to deal with the triangle singularities is done in Ref.~\cite{guo}, with a different method to perform the integrals and an easy and intuitive rule to determine where the singularities appear.

Recent examples of processes where the triangle singularities are relevant can be seen in the $\eta(1405) \rightarrow \pi\,a_0(980)$ and $\eta(1405) \rightarrow \pi\,f_0(980)$ \cite{wu2,wu1,wu3}. The latter process is isospin forbidden and results largely enhanced due to a triangle singularity involving $\eta(1405) \rightarrow K^*\bar K$ followed by $K^*\rightarrow K\,\pi$ and fusion of $K\,\bar K$ to give the $f_0(980)$. A more recent example can be seen in the COMPASS collaboration \cite{compass}, associated to a new resonance ``$a_1(1420)$", which, as hinted in Ref.~\cite{qzhao} and proved in Refs.~\cite{mikha,aceti}, comes from the $\pi f_0(980)$ decay of the $a_1(1260)$, via a triangle singularity proceeding through $a_1\rightarrow K^*\bar K$, $K^*\rightarrow K\pi$ and $K\bar K\rightarrow f_0(980)$. Related to this is the recent interpretation of the $f_1(1420)$ resonance as a decay mode of the $f_1(1285)$ into $\pi a_0(980)$ and $K^*\bar K$ \cite{vinicius}. Another interesting example is the role played by a triangle singularity in the $\gamma p\rightarrow K^+\Lambda(1405)$ reaction \cite{wang}. The process $\gamma p\rightarrow K^*\Sigma$, $K^*\rightarrow K\pi$, $\Sigma\pi\rightarrow \Lambda(1405)$ leads to a peak in the cross section around $\sqrt{s}=2120$ MeV that solved a standing problem in that reaction.

Similarly, the $f_2$(1810) is also explained as a consequence of the $f_2(1650)\rightarrow K^*\bar K^*$, $K^*\rightarrow \pi K$ and $K\bar K^*$ merging into the $a_1$(1260) \cite{geng}. Other examples can be found in Refs.~\cite{hanhart,achasov,Lorenz,adam1,adam2}. Renewed interest in the triangular singularities came from the suggestion that the narrow peak of the $J/\psi\,p$ invariant mass at 4450 MeV seen by the LHCb collaboration \cite{penta-ex,chinese}, and interpreted there as a pentaquark state, could be due to a triangle singularity with $\Lambda_b\rightarrow \Lambda(1890) \chi_{c1}$, $\Lambda(1890)\rightarrow \bar K p$, $p\chi_{c1}\rightarrow J/\psi p$ \cite{ulfguo,liu}. However, as shown in Ref.~\cite{guo}, for the preferred experimental quantum number of this peak, $3/2^-$, $5/2^+$, the $\chi_{c1} p\rightarrow J/\psi p$ proceeds with $\chi_{c1} p$ in $p$-wave or $d$-wave and the $\chi_{c1} p$ threshold is exactly 4450 MeV, hence, this amplitude vanishes there on shell and the suggested process cannot be responsible for the observed peak.

In the present work we will show that the process of the $N^*(1875)\rightarrow \Sigma^*K$, $\Sigma^*\rightarrow \pi\Lambda$, $\Lambda K\rightarrow N^*(1535)$, develops a triangle singularity precisely at the same mass of resonance and reinforces it. The other interesting finding of this work is that there can also be triangle mechanisms, which, without developing a singularity, can be very important. This is actually the case with $N^*(1875)\rightarrow \Delta\pi$, $\Delta\rightarrow \pi N$, $\pi\pi\rightarrow f_0(500)$. We shall see that because of the large strength of all the couplings involved, this process becomes even more important than the $N^*(1875)\rightarrow \pi N^*(1535)$ and leads to a sizeable partial decay width $N^*(1875)\rightarrow N\sigma (f_0(500))$.

\section{Formalism}
\subsection{Brief review of the pseudoscalar meson-baryon decuplet interaction}
Following Ref. \cite{sarkar}, the sector with $S=0$, $I=\frac{1}{2}$ is reached with the channels $\Delta\pi$, $\Sigma^*K$. In $s$-wave the interaction leads to $J^P=3/2^-$ states. The interaction is given by
\begin{equation}\label{WT-int}
V_{ij} = -C_{ij}\,\frac{1}{4f^2}\,(k^0 + k'^0),
\end{equation}
where $k^0,~k'^0$ are energies of the initial and final mesons respectively and the coefficients $C_{ij}$ are given in Table \ref{cij}.
\begin{table}[b]
\begin{center}
\begin{tabular}{c| c c }
\hline\hline
 & $\Delta\pi$ & $\Sigma^*K$ \\
 \hline
$\Delta\pi$ & 5 & 2 \\
$\Sigma^*K$  & 2 & 2\\
\hline\hline
\end{tabular}
\caption{\label{cij}The $C_{ij}$ coefficients of the Eq. (\ref{WT-int}).}
\end{center}
\end{table}
The scattering matrix is given via the Bethe-Salpeter equation in the matrix form by
\begin{eqnarray}\label{BS-eqn}
T = \big[ 1 - VG\big]^{-1} V,
\end{eqnarray}
where $G$ is the ordinary meson-baryon loop function. The $\Delta\pi\rightarrow \Delta\pi$ amplitude develops a strong peak around 1500 MeV that was associated in Ref.~\cite{sarkar} to the $N^*$(1520) resonance. By contrast, this amplitude is very small around 1875 MeV, as a consequence of interference of terms, and it is the $\Sigma^*K\rightarrow \Sigma^*K$ amplitude the one that shows as a clear peak around 1875 MeV. In the next subsection we shall include the $N^*(1535)\pi$ and $N f_0(500)$ channels.

\subsection{The $N^*$(1535)$\pi$ channel}
\begin{figure}[tbp!]\centering
\includegraphics[scale=0.8]{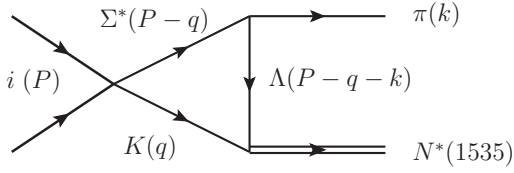}
\caption{The triangle diagram for the $i(\Delta \pi,\,\Sigma^* K) \to N^*(1535)\pi$ transition. In brackets, the momenta of the lines.
\label{fig:Fig1}}
\end{figure}

\begin{figure*}[tbh]\centering
\includegraphics[scale=0.7]{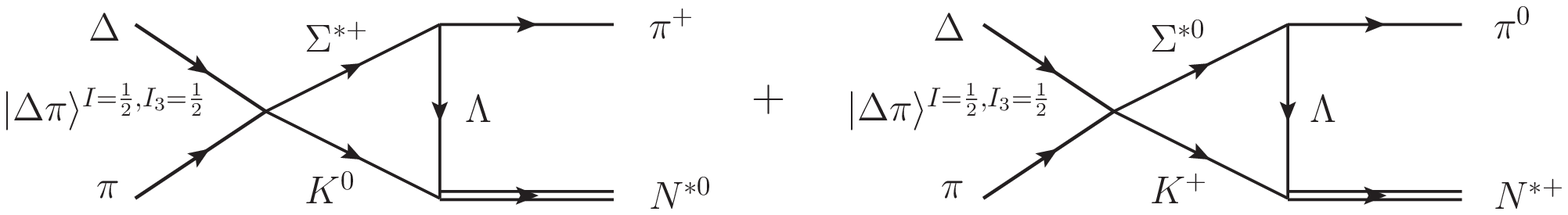}
\caption{Different diagrams leading to $\Delta \pi \to N^*\pi \;(I=\frac{1}{2}, I_3=\frac{1}{2})$.
\label{fig:Fig2}}
\end{figure*}

\begin{figure*}[tbh]\centering
\includegraphics[scale=0.7]{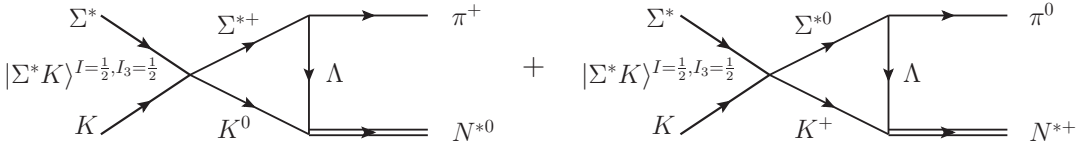}
\caption{Different diagrams leading to $\Sigma^* K \to N^*\pi \;(I=\frac{1}{2}, I_3=\frac{1}{2})$.
\label{fig:Fig3}}
\end{figure*}

We shall look into the diagram of Fig. \ref{fig:Fig1}, where the state $i$ stands for $\Delta\pi$ and $\Sigma^*K$. By looking at equation (18) of Ref.~\cite{guo}, the relationship
\begin{eqnarray}\label{TS-relation}
q_{{\rm on}\,+} - q_{a\,-} = 0 \,,
\end{eqnarray}
where $q_{on\,+}$ and $q_{a\,-}$ are defined by easy analytical expression in Ref.~\cite{guo}, shows the energy $\sqrt{s}$ at which a triangle singularity appears. One must check Eq. \eqref{TS-relation} for a mass of the $N^*$(1535) bigger than $m_\Lambda + m_K$. At a mass about 1615 MeV, which is in the range of the $N^*$(1535) mass considering the width (150 MeV), Eq. \eqref{TS-relation} shows a solution at 1878 MeV. But a peak in the amplitude develops for smaller $N^*$ masses within the range of the $N^*$(1535) spectral function, which we shall take into account in the evaluation of the diagram of Fig. \ref{fig:Fig1}.
Since we are looking into the states with isospin $I=\frac{1}{2}$ we must consider in detail, the different charge combinations that enter the evaluation of Fig. \ref{fig:Fig1}. This is shown in Figs. \ref{fig:Fig2} and \ref{fig:Fig3}, where the state $i$ is $\Delta\pi$ or $\Sigma^*K$, respectively.

We must project all of them into $I=\frac{1}{2}$ and sum the diagrams. For this, let us recall the sign convention in Ref.~\cite{sarkar}, $|\pi^+\rangle = -|1,1\rangle$, of isospin. Hence we have
\begin{eqnarray}\label{isospindecom}\allowdisplaybreaks
&&\!\!\left|\Delta\pi, I=\textstyle{\frac{1}{2}}, I_3=\textstyle{\frac{1}{2}} \right\rangle
\nonumber\\
&=& \!\!\sqrt{\frac{1}{2}}\left| \Delta^{++}\pi^-\right\rangle
- \sqrt{\frac{1}{3}}\left| \Delta^{+}\pi^0\right\rangle -\sqrt{\frac{1}{6}}\left| \Delta^{0}\pi^+\right\rangle,
\nonumber\\
&&\!\!\left|N^*(1535)\pi, I=\textstyle{\frac{1}{2}}, I_3=\textstyle{\frac{1}{2}} \right\rangle
\nonumber\\
& =& \!\!\sqrt{\frac{2}{3}}\left| N^{*\,0}\pi^+\right\rangle
+ \sqrt{\frac{1}{3}}\left| N^{*\,+}\pi^0\right\rangle,\\
&&\!\!\left|\Sigma^*K, I=\textstyle{\frac{1}{2}}, I_3=\textstyle{\frac{1}{2}} \right\rangle
 = \!\sqrt{\frac{2}{3}}\left| \Sigma^{*\,+}K^0\right\rangle -\sqrt{\frac{1}{3}}\left| \Sigma^{*\,0}K^+\right\rangle. \nonumber
\end{eqnarray}
We need the $\Sigma^*\rightarrow \Lambda\pi$ coupling and the $K\Lambda\rightarrow N^*(1535)$ coupling. The first one is of the type
\begin{eqnarray}\label{eq:tSigpiLam}
-i t_{\Sigma^*,\pi\Lambda} = g_{\Sigma^*,\pi\Lambda}\vec S\cdot \vec k,
\end{eqnarray}
where $\vec S$ is the spin transition operator from $3/2$ to $1/2$. The width for $\Sigma^*\rightarrow \pi\Lambda$ is given by $( \vec k \equiv \vec p_\pi)$
\begin{eqnarray}
\Gamma = \frac{2M_\Lambda 2M_{\Sigma^*}}{8\pi}\frac{1}{M_{\Sigma^*}^2}\;\overline\sum \sum |t|^2 \;|\vec k|,
\end{eqnarray}
with
\allowdisplaybreaks
\begin{eqnarray}
&&\overline\sum \sum |t|^2
\nonumber
\\
&&\quad = \frac{1}{4}g_{\Sigma^*,\pi\Lambda}^2\sum_M\sum_m\langle m|S_i|M\rangle\langle M|S_j^\dagger|m\rangle k_i k_j
\nonumber\\
&&\quad = \frac{1}{4}g_{\Sigma^*,\pi\Lambda}^2\sum_m\langle m|{\textstyle{\frac{2}{3}}}\delta_{ij} - {\textstyle{\frac{i}{3}}}\epsilon_{ijl}\sigma_l|m\rangle k_i k_j
\nonumber\\
&&\quad = \frac{1}{3}\;g_{\Sigma^*,\pi\Lambda}^2\; \vec k^2 .
\end{eqnarray}
Hence,
\begin{eqnarray}
\Gamma = \frac{1}{2\pi}\frac{M_\Lambda}{M_{\Sigma^*}}\;\frac{1}{3}g_{\Sigma^*,\pi\Lambda}^2 \;k^3,
\end{eqnarray}
and using the experimental value for the $\Sigma^*\rightarrow \pi\Lambda$ width we obtain
\begin{eqnarray}
g_{\Sigma^*,\pi\Lambda} = 0.0090~{\text MeV}^{-1}.
\end{eqnarray}
The coupling of $N^*$ to $K\Lambda$ we get from Ref.~\cite{inoue}, where the chiral unitary approach has been used to obtain $\pi N$ scattering in the region of the $N^*$(1535). One has
\begin{equation}
g_{N^*,K\Lambda} = -1.28.
\end{equation}
With these ingredients we can already evaluate the triangle diagrams of Figs. \ref{fig:Fig2} and \ref{fig:Fig3}. Considering the isospin coefficients, the sum of the diagrams in Fig. \ref{fig:Fig2}, for $I=\frac{1}{2}$, $I_3=\frac{1}{2}$, up to the propagators, is given by
\begin{eqnarray}\label{VNstpi}
&&(-i)\sqrt{\frac{2}{3}}\;V_{\Delta\pi,\Sigma^*K}^{(I=1/2)}\;(-1)g_{\Sigma^*,\pi\Lambda}\;(-i)g_{N^*,K\Lambda} \;\sqrt{\frac23}
\nonumber\\
&&+ (-i)(-1)\sqrt{\frac{1}{3}}\;V_{\Delta\pi,\Sigma^*K}^{(I=1/2)}\;g_{\Sigma^*,\pi\Lambda}\;(-i)g_{N^*,K\Lambda} \;\sqrt{\frac13}
\nonumber\\
&=& V_{\Delta\pi,\Sigma^*K}^{(I=1/2)} \;g_{\Sigma^*,\pi\Lambda}\;g_{N^*,K\Lambda},
\end{eqnarray}
and the full contribution of the loop is given by
\begin{widetext}
\begin{eqnarray}\label{tDelpi-Nstpi}
-i t_{\Delta\pi,\pi N^*} \!\!&=& \!\!V_{\Delta\pi,\Sigma^*K}^{(I=1/2)} \;g_{\Sigma^*,\pi\Lambda} \;g_{N^*,K\Lambda}
(\vec S\cdot \vec k) \;2M_\Lambda 2M_{\Sigma^*}
\!\int \!\!\frac{{\rm d}^4q}{(2\pi)^4}\frac{i}{(P-q)^2 - M_{\Sigma^*}^2 + i\epsilon}\frac{i}{(P-q-k)^2 - M_\Lambda^2 + i\epsilon}\frac{i}{q^2-m_K^2 +i\epsilon}
\nonumber\\
&\equiv& V_{\Delta\pi,\Sigma^*K}^{(I=1/2)}\;g_{\Sigma^*,\pi\Lambda}\;g_{N^*,K\Lambda}
(\vec S\cdot \vec k) \;2M_\Lambda 2M_{\Sigma^*}\; t_T,
\end{eqnarray}
\end{widetext}
where the last equation defines the triangle integral $t_T$. The factors $2M_\Lambda$, $2M_{\Sigma^*}$ are consequence of using the Mandl and Shaw normalization for the Fermion fields \cite{mandl}. This integral is performed by doing analytically the $q^0$ integration, and we obtain \cite{dias,guo}
\begin{widetext}
\begin{eqnarray}\label{t_T}
t_T &=& \int\frac{d^3q}{(2\pi)^3}\frac{1}{8\omega_K E_{\Sigma^*} E_\Lambda}\frac{1}{k^0 - E_\Lambda - E_{\Sigma^*}}\frac{1}{P^0 + \omega_K + E_\Lambda - k^0}\frac{1}{P^0 - \omega_K - E_\Lambda - k^0 + i\epsilon}\frac{1}{P^0 - E_{\Sigma^*} - \omega_{K} + i\epsilon}
\nonumber\\
&& \times \big\{ 2P^0\omega_K + 2k^0E_\Lambda -2(\omega_K + E_\Lambda)(\omega_K + E_\Lambda + E_{\Sigma^*})\big\},
\end{eqnarray}
where
\begin{eqnarray}
\omega_K &=& \sqrt{m_K^2 + \vec q^2},~~~~~~~~\quad E_{\Sigma^*} = \sqrt{M_{\Sigma^*}^2 + \vec q^2} +\frac{i\Gamma_{\Sigma^*}}{2}, ~~~~~\qquad E_\Lambda = \sqrt{M_\Lambda^2 + (\vec q + \vec k)^2},
\nonumber\\
k^0 &=&\frac{s+m_\pi^2-M_{N^*}^2}{2\sqrt{s}},\qquad |\vec k| = \frac{\lambda^{\frac{1}{2}}(s,m_\pi^2,M_{N^*}^2)}{2\sqrt{s}},
\qquad \lambda(x,y,z) = x^2 + y^2 + z^2 - 2xy - 2yz -2xz\,.
\end{eqnarray}
\end{widetext}
The case of the transition $\Sigma^*K\rightarrow  N^*\pi$ in Fig. \ref{fig:Fig2} proceeds in an identical way, and the only difference with respect to the results of Eq. (\ref{tDelpi-Nstpi}) is that we must substitute $V_{\Delta\pi,\Sigma^*K}^{(I=1/2)}$ by $V_{\Sigma^*K,\Sigma^*K}^{(I=1/2)}$.

We must note that originally the $\vec S\cdot \vec k$ operator appeared for the $\Sigma^*\rightarrow \pi\Lambda$ transition, but upon sum of the intermediate $\Sigma^*$ spin components it becomes now in Eq. (\ref{tDelpi-Nstpi}) the spin transition operator from $\Delta$ to $N^*$ because the $s$-wave potentials $V_{\Delta\pi,\Sigma^*K}^{(I=1/2)}$ and $V_{,\Sigma^*K,\Sigma^*K}^{(I=1/2)}$ are independent of the $\Delta$ and $\Sigma^*$ spins.

Neglecting the width of the $\Sigma^*$ in Eq. (\ref{t_T}), the integrand in $t_T$ will have poles when
\begin{eqnarray}
P^0 - k^0 - \omega_K - E_\Lambda = 0 \; {\text{and}}\; P^0 - E_{\Sigma^*} - \omega_K =0.
\end{eqnarray}
In principle, in the integral they will give rise to imaginary parts and principal values, via the $i\epsilon$. However, the cancellations in the principal values will not occur when we are at the extremes of $\cos\theta$ ($\hat k\cdot\hat q$) when $\cos\theta = \pm 1$. Then a singularity can appear in the integral, triangle singularity (which, occurs for $\cos\theta=-1$ \cite{guo}), which however, is rendered finite when the width of the $\Sigma^*$ is explicitly considered \cite{guo}. The integral in the $t_T$ is then convergent, but we perform a cut off in $q$ in the rest mass of the $N^*$, when the chiral unitary approach is done, and we use $q_{\rm max}^{\rm cm}\equiv 1000 ~{\rm MeV}$, suited to the results of Ref.~\cite{inoue}.

\begin{figure}[b!]\centering
\includegraphics[scale=0.5]{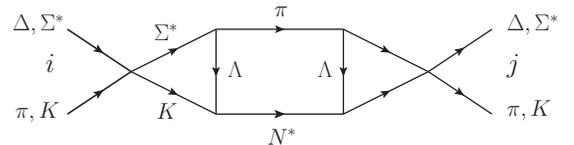}
\caption{Transition diagram from $\Delta \pi (\Sigma^* K ) \to \Delta \pi (\Sigma^* K )$ via the intermediate $N^* \pi$.
\label{fig:Fig4}}
\end{figure}

We would like to include now the $\pi N^*$ into the coupled channels, together with $\Delta\pi$ and $\Sigma^*K$. However we can see that while the interaction between $\Delta\pi$ and $\Sigma^*K$ proceeds via $s$-wave, the transition $\Delta\pi\rightarrow \pi N^*$ proceeds via $p$-wave with the $\vec S\cdot \vec k$ operator. This is a consequence from the transition of a $\Delta(3/2^+)$ to $N^*(1/2^-)$ which requires change of parity. Yet, it is possible to mix the channels via an effective $s$-wave potential, as done in Refs.~\cite{garzon,vectorrev,uchino1,uchino2}. In order to define this effective potential we look at the diagram of Fig. \ref{fig:Fig4}, which makes transitions from $\Delta\pi(\Sigma^*K)\rightarrow \Delta\pi(\Sigma^*K)$ via an intermediate $\pi N^*$ state.
We can write for the transition amplitude
\begin{equation}\label{tij-Nstpi}
-i t_{ij} = -i t_{i,\pi N^*} \;i G_{\pi N^*}\; (-i) t_{\pi N^*,j}.
\end{equation}
In the chiral unitary approach the transition potentials are evaluated for the external lines on shell and we wish to do the same with the new channel $\pi N^*$. For this purpose, we take the imaginary part of $G_{\pi N^*}$ in Eq. (\ref{tij-Nstpi}), which places $\pi N^*$ on shell. Considering that in the $\pi N^*\rightarrow \Delta\pi(\Sigma^*K)$ transition the pion momentum is in going instead of outgoing as in $\Delta\pi(\Sigma^*K)\rightarrow \pi N^*$, we have
\begin{eqnarray}\label{Imtij}
{\rm Im}t_{i,j} &=& -V_{j,\Sigma^* K} \;V_{i,\Sigma^*K}\;\big( g_{\Sigma^*,\pi\Lambda}\;g_{N^*,K\Lambda}\;2M_\Lambda2M_{\Sigma^*}\big)^2
\nonumber\\
&&\times(-)(\vec S^\dagger\cdot\vec k)\; (\vec S\cdot\vec k)\; t_T\, t_T^* \;{\rm Im}G_{\pi N^*}.\nonumber
\end{eqnarray}
Now we have\footnote{Note the order of $\vec S^\dagger\,,\,\vec S$ and the sum over $m$, the spin of the $1/2$ baryon.}
\allowdisplaybreaks
\begin{eqnarray}\label{s-d wave}
&&\sum_m\langle M'|\vec S^\dagger\cdot\vec k| m \rangle\langle m|\vec S\cdot \vec k|M\rangle
= \frac{1}{3}\vec k^2\delta_{MM'}
\nonumber \\
&-& \frac{1}{3}\vec q^{\;2}\sqrt{4\pi}\mathcal{C}(\frac{3}{2}\,2\,\frac{3}{2};\, M',M\!\!-\!\!M')Y_{2,M\!-\!M'}(\hat q),
\end{eqnarray}
which indicate that we can have transitions in $s$- and $d$-waves. But we are only interested in the $s$-wave and hence we keep the $\frac13 \vec k^2\delta_{MM'}$ factor in Eq. (\ref{s-d wave}). Thus, effectively we can take
\begin{eqnarray}\label{Imttildeij}
{\rm Im}\tilde t_{i,j} &=& V_{j,\Sigma^* K} \;V_{i,\Sigma^*K} \;\big( g_{\Sigma^*,\pi\Lambda}\;g_{N^*,K\Lambda}\;2M_\Lambda2M_{\Sigma^*}\big)^2
\nonumber\\
&&\times t_T t_T^* \;\frac13 \vec k^2 \;{\rm Im}G_{\pi N^*}
\end{eqnarray}
with
\begin{eqnarray}\label{ImG}
{\rm Im}G_{\pi N^*} = -\frac{1}{4\pi}\frac{M_{N^*}}{\sqrt s}|\vec k|, ~~~ |\vec k|=\frac{\lambda^{\frac{1}{2}}(s,m_\pi^2,M_{N^*}^2)}{2\sqrt{s}}.~~
\end{eqnarray}
However, since the triangle singularity is sensitive to the external masses and the $N^*$ has a width of 150 MeV, we make a convolution of Eq. (\ref{Imttildeij}) with the spectral function of the $N^*$, such that
\begin{eqnarray}\label{intImtij}
{\rm Im}\tilde t_{i,j}\!\!\! &=& \!\!\! V_{j,\Sigma^* K} \;V_{i,\Sigma^*K}\big( g_{\Sigma^*,\pi\Lambda}\,g_{N^*,K\Lambda}\,2M_\Lambda2M_{\Sigma^*}\big)^2
\nonumber\\
\!\!\!\!\!&\times&\!\!\!\!\frac{1}{N_{N^*}}\!\!\!\int\! \!\!{\rm d}\tilde m \frac{1}{3} \tilde k^2 S_{N^*}(\tilde m){\rm Im}G_{\pi N^*}(s,\tilde m) |t_T(s,\tilde m)|^2,~~~~~
\end{eqnarray}
where ${\rm Im}G_{\pi N^*}(s,\tilde m)$, $\tilde k$ and $t_T$ are obtained substituting $M_{N^*}\rightarrow \tilde m$ in Eqs. (\ref{ImG}) and (\ref{t_T}). The spectral function of the $N^*$ is given by
\begin{eqnarray}\label{eq:specFun}
S_{N^*}(\tilde m) = -\frac{1}{\pi}{\rm Im}\frac{1}{\tilde m - M_{N^*} + i\frac{\Gamma_{N^*}}{2}},
\end{eqnarray}
and the factor $N_{N^*}$ in Eq.~(\ref{intImtij}) is put for normalization
\begin{eqnarray}\label{normal}
N_{N^*}=\int S_{N^*}(\tilde m) d\tilde m .
\end{eqnarray}
The limits of $\tilde m$ in Eqs. (\ref{intImtij}) and (\ref{normal}) are taken from $M_{N^*} - \alpha \Gamma_{N^*}$ to $M_{N^*} + \alpha \Gamma_{N^*}$ with $\alpha$ around one or two. The $\tilde m$ dependence in Eq. (\ref{intImtij}) does not effect $V_{j,\Sigma^* K}V_{i,\Sigma^*K}$, hence, we can define
\begin{eqnarray}
{\rm Im}\tilde t &=& \frac{1}{N}\int d\tilde m
\nonumber\\
&&\times\frac13 \tilde k^2 S_{N^*}(\tilde m){\rm Im}G_{\pi N^*}(s,\tilde m) |t_T(s,\tilde m) |^2
\end{eqnarray}
and a function $\tilde V$ such that
\begin{eqnarray}
{\rm Im}\tilde t = \tilde V\; {\rm Im}G_{\pi N^*}\tilde V.
\end{eqnarray}
then we can construct an effective $s$-wave transition potential
\begin{eqnarray}\label{eff-s-wave}
\tilde V_{i,\pi N^*} = V_{i,\Sigma^*K}^{(I=1/2)}\;g_{\Sigma^*,\pi\Lambda}\;g_{N^*,K\Lambda}\;2M_\Lambda2M_{\Sigma^*}\;\tilde V
\end{eqnarray}
such that
\begin{eqnarray}
\tilde V_{i,\pi N^*}\; {\rm Im}G_{\pi N^*}(M_{N^*})\;\tilde V_{j,\pi N^*} = {\rm Im}\tilde t_{ij}.
\end{eqnarray}
This means that using $\tilde V_{i,\pi N^*}$ in coupled channels with the extra $\pi N^*$ channel we can effectively incorporate the mechanism of Fig. \ref{fig:Fig4} and when the resonance shows up in the amplitudes we can evaluate the coupling of the resonance to the $\pi N^*$ channel and then the partial decay width into this channel. We will have now a new $V$ matrix, containing the $\Delta\pi(\Sigma^*K)\rightarrow \Delta\pi(\Sigma^*K)$ of Eq. (\ref{WT-int}) plus the $\Delta\pi(\Sigma^*K)\rightarrow \pi N^*$ transition of Eq. (\ref{eff-s-wave}). We do not include a direct $\pi N^*\rightarrow \pi N^*$ transition assuming such transition would occur via the $\Delta\pi(\Sigma^*K)$ intermediate states involving the $\Delta\pi(\Sigma^*K)\rightarrow \pi N^*$ transition which contains the triangle diagram.

In order to take into account the $\Delta$ and $\Sigma^*$ widths in the $G$ functions of Eq.~(\ref{BS-eqn}) we also do a convolution, as done in Ref.~\cite{sarkar}, with the spectral function of the baryons $S_B(\tilde M)$,
\begin{eqnarray}
G\rightarrow \tilde G = \frac{1}{N}\int d\tilde M\, G(\tilde M) S_B(\tilde M).
\end{eqnarray}

\subsection{The $Nf_0(500)(\sigma)$ channel}\label{subsec:Nf0_channel}
\begin{figure}[h]\centering
\includegraphics[scale=0.6]{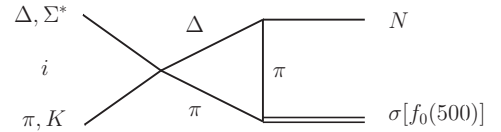}
\caption{Triangle diagram for the transition of $i(\Delta \pi,\,\Sigma^* K)$ to $N\sigma$.
\label{fig:Fig5}}
\end{figure}

\begin{figure*}[htp!]\centering
\includegraphics[scale=0.57]{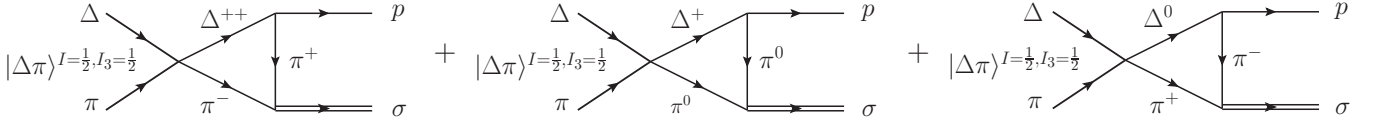}
\caption{Diagrams for the $|\Delta\pi, I=\frac{1}{2}, I_3=\frac{1}{2}\rangle$ transition to $p\sigma$.
\label{fig:Fig6}}
\end{figure*}

\begin{figure*}[htp]\centering
\includegraphics[scale=0.57]{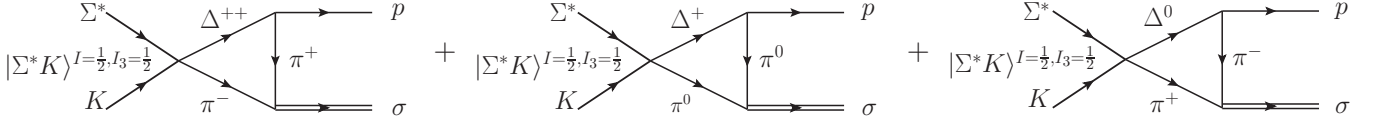}
\caption{Diagrams for the $|\Sigma^* K, I=\frac{1}{2}, I_3=\frac{1}{2}\rangle$ transition to $p\sigma$.
\label{fig:Fig7}}
\end{figure*}

We can now consider a triangle diagram which involves $\Delta \pi$ instead of $\Sigma^* K$ in the intermediate states. This is depicted in Fig. \ref{fig:Fig5}.
The states $\Delta \pi,\,\Sigma^* K$ can now make transition to the $\Delta \pi$, the $\Delta$ decays into $\pi N$ and then the two pions fuse to give the $f_0(500)(\sigma)$. The first thing one must check is if this diagram can develop a singularity at some energy $\sqrt{s}$. Application of Eq. (\ref{TS-relation}) immediately tells us that this is not the case, and $q_{{\rm on}+}-q_{a-}$ does not vanish for any energy of the original system. However, we have now other elements to make this mechanism particularly relevant. First we can have now $\Delta \pi \to \Delta \pi$ transitions that have a weight of a factor 5 (see table \ref{cij}) instead of 2, as we had before. Second, the $\Delta \to \pi N$ coupling is very large and so is the coupling of the $\pi\pi$ to the $\sigma$. The evaluation of the $\Delta \pi (\Sigma^* K) \to N\sigma$ transition proceeds in an analogous way to the previous subsection. First, in analogy to Fig. \ref{fig:Fig2} and Fig. \ref{fig:Fig3} we have now Fig. \ref{fig:Fig6} and Fig. \ref{fig:Fig7}.

We need now the state
\begin{equation}\label{eq:pipi}
  |\pi\pi, I=0 \rangle =-\frac{1}{\sqrt{3}} (\pi^+ \pi^- +\pi^-\pi^+ +\pi^0 \pi^0)
\end{equation}
and the $\Delta \to N\pi$ coupling, similar to Eq. \eqref{eq:tSigpiLam}
\begin{equation}\label{eq:delNpi}
  -it=\frac{f_{\pi N \Delta}}{m_\pi} \vec{S}\cdot \vec{p}_\pi \mathcal{C}(i),
\end{equation}
with $\mathcal{C}(i)$ the corresponding isospin Clebsch-Gordan coefficient,
\begin{align}\label{eq:Ci}
   & \mathcal{C}(i)
  = \left\{
  \begin{array}{ll}
    -1, & {\rm for~} \Delta^{++}\to p\pi^+ \\[3mm]
   \sqrt{\frac{2}{3}}, & {\rm for~} \Delta^{+}\to p\pi^0\\[3mm]
   \sqrt{\frac{1}{3}}, & {\rm for~} \Delta^{0}\to p\pi^-\\
    \end{array}
   \right.
\end{align}
The coupling $f_{\pi N \Delta}$, taken to obtain the $\Delta$ width, is given by
\begin{equation}\label{eq:fpiNDel}
  f_{\pi N \Delta}=2.2
\end{equation}
corresponding to $f^2_{\pi N N}/4\pi =0.38$, very close to standard value used in pion physics, 0.36 \cite{ericson}.
The isospin combination of vertices corresponding to Eq. (\ref{VNstpi}) for Fig. \ref{fig:Fig6} is now given, taking into account Eq. (\ref{isospindecom}), by
\begin{eqnarray}\label{eq:ContriFig7}
   && -i\, V^{(I=1/2)}_{{\Delta \pi, \Delta \pi}}\, (-) \sqrt{\frac{1}{2}}\, \frac{f_{\pi N\Delta}}{m_\pi}\;  (-) \frac{1}{\sqrt{3}}\, (-i) g_{\sigma, \pi\pi}\nonumber\\
  & & -i\, V^{(I=1/2)}_{{\Delta \pi, \Delta \pi}}\, (-) \sqrt{\frac{1}{3}}\, \sqrt{\frac{2}{3}}\;\frac{f_{\pi N\Delta}}{m_\pi}\;  (-) \frac{1}{\sqrt{3}}\, (-i) g_{\sigma, \pi\pi}\nonumber\\
  && -i\, V^{(I=1/2)}_{{\Delta \pi, \Delta \pi}}\, (-) \sqrt{\frac{1}{6}}\,\sqrt{\frac{1}{3}}\, \frac{f_{\pi N\Delta}}{m_\pi}\;  (-) \frac{1}{\sqrt{3}}\, (-i) g_{\sigma, \pi\pi} \nonumber\\
  &=& -\sqrt{\frac{2}{3}} \, V^{(I=1/2)}_{{\Delta \pi, \Delta \pi}}\; \frac{f_{\pi N\Delta}}{m_\pi}\; g_{\sigma, \pi\pi}.
\end{eqnarray}

For the coupling of the $\sigma$ to $\pi\pi$ obtained from the unitary matrix, and unitary normalization ($\frac{1}{\sqrt{2}}$ extra in the wave function of $\pi\pi$ as identical particles) we take
\begin{equation}\label{eq:gsigmaNN}
  g'_{\sigma, \pi\pi}=3.6~{\rm GeV},
\end{equation}
where we have taken an average between the results of the chiral unitary approach \cite{nsd} and different results using an analysis of data implementing Roy equations \cite{colangelo,robert} (see table 4 of Ref.~\cite{reviewpel}). With the good normalization to be used in Eq. (\ref{eq:ContriFig7}), we have
\begin{equation}\label{eq:gsigmapipi}
  g_{\sigma, \pi\pi}=\sqrt{2}\times 3.6~{\rm GeV}.
\end{equation}

Following the argumentation of Eq. (\ref{tDelpi-Nstpi}), we obtain now
\begin{equation}\label{eq:tdelpiNsig}
  -it_{\Delta \pi, N\sigma}=-V^{(I=1/2)}_{{\Delta \pi, \Delta \pi}}\, \sqrt{\frac{2}{3}}\,\frac{f_{\pi N\Delta}}{m_\pi}\; g_{\sigma, \pi\pi}\, (\vec{S}\cdot \vec{k})\; 2M_{\Delta}\, t'_T,
\end{equation}
with $\vec k$ the nucleon momentum, where $t'_T$ is obtained from Eq. (\ref{t_T}) by simply changing the masses of the intermediate particles $\Sigma^* \to \Delta, K \to \pi, \Lambda \to \pi$ and multiplying the integrand by $\left( 1+ \frac{\vec{q}\cdot \vec{k}}{|\vec{k}|^2}\right)$. The reason for this latter factor is that before in Eq. (\ref{tDelpi-Nstpi}) the factor $(\vec{S}\cdot \vec{k})$, factorized outside the integral. Here we have $\vec{S}\cdot \vec{p}_\pi \equiv \vec{S}\cdot (-\vec q-\vec k)$ and
\begin{equation*}
  \int d^3q \;q_i \cdots \equiv k_i \int d^3q \,\frac{\vec q \cdot \vec k}{\vec k^2},
\end{equation*}
since $\vec q$ is an integration variable and $\vec k$ is the only vector in the integrand which is not integrated.

For the transition of $\Sigma^* K \to N\sigma$, we will have the same expression as in Eq. \eqref{eq:tdelpiNsig} changing $V^{(I=1/2)}_{{\Delta \pi, \Delta \pi}}$ to $V^{(I=1/2)}_{{\Sigma^* K, \Delta \pi}}$

Finally, in analogy to Eq. (\ref{eff-s-wave}) we will now have the effective transition potential
\begin{eqnarray}\label{eq:V_effective4}
\tilde{V}_{i, N\sigma}&=& V^{I=1/2}_{i, \Delta \pi}  \sqrt{\frac{2}{3}}\,\frac{f_{\pi N\Delta}}{m_\pi} \; g_{\sigma, \pi \pi} \, 2M_\Delta  \; \tilde{V}'_{N\sigma}(s),~
\end{eqnarray}
where $\tilde{V}'_{N\sigma}$ is defined such that
\begin{equation}\label{eq:V_effective2}
  {\rm Im} \tilde{t}'(s)=\tilde{V}'_{N\sigma}(s)\, {\rm Im}G_{N\sigma}(s,m_\sigma)\, \tilde{V}'_{N\sigma}(s),
\end{equation}
with
\begin{eqnarray}\label{eq:Imt'2}
 {\rm Im}\tilde{t}'(s)\!\!&=&\!\!\frac{1}{N_\sigma}\, \int d\tilde{m}^2_\sigma \nonumber \\
 &\times& \!\!\!\frac{1}{3} \tilde{k}^2 \,
 {\rm Im} G_{N\sigma}(s,\tilde{m}_\sigma)\, S_{\sigma}(\tilde{m}^2_\sigma)\, |t'_T(s,\tilde{m}_\sigma)|^2,~~~~
\end{eqnarray}
with the $\sigma$ spectral function
\begin{equation}\label{eq:Ssigma}
  S_{\sigma}(\tilde{m}^2_\sigma)=-\frac{1}{\pi}\, {\rm Im}\left[ \frac{1}{\tilde{m}^2_\sigma-m^2_\sigma+im_\sigma\Gamma_\sigma} \right],
\end{equation}

\begin{equation}\label{eq:N}
 N_\sigma= \int S_{\sigma}(\tilde{m}^2_\sigma) \,d\tilde{m}^2_\sigma,
\end{equation}
\begin{equation*}
  {\rm Im} G_{N\sigma}(s, {m}_\sigma)=-\frac{1}{4\pi}\,\frac{M_N}{\sqrt{s}}\,k(s, m_\sigma),
\end{equation*}
\begin{equation*}
  k(s, m_\sigma)=\frac{\lambda^{1/2}(s,M^2_N, m^2_\sigma)}{2\sqrt{s}},
\end{equation*}
and ${\rm Im} G_{N\sigma}(s, \tilde{m}_\sigma)$, $\tilde{k}(s, \tilde{m}_\sigma)$ given by the same expressions changing $m_\sigma$ to $\tilde{m}_\sigma$. For $m_\sigma$ and $\Gamma_\sigma$ we take values from Ref.~\cite{reviewpel}
\begin{equation*}
  m_\sigma = 445~ {\rm MeV}, ~~~~~~\frac{\Gamma_\sigma}{2}=275 ~{\rm MeV}.
\end{equation*}

Now $\tilde{V}_{i, N\sigma}$ of Eq.~\eqref{eq:V_effective4} provides transitions from $\Delta\pi$($\Sigma K$) to $N\sigma$. As before, we introduce the $N\sigma$ channel into the coupled channels and have now a $4\times 4$ matrix for $V$, allowing the $\Delta\pi \to N\sigma$, $\Sigma^* K\to N\sigma$ transitions and neglecting direct transition $N\sigma \to N\sigma$ and $N\sigma \to N^* \pi$. For cutoff in the integral of ${\mathrm{d}^3q}$ in $t'_T$ we take now $q_{\rm max}=700~{\rm MeV}$, suited for the study of $\pi\pi$ interaction \cite{npa,weihong}.

\subsection{Couplings and partial decay widths}

In order to obtain the couplings we look at the amplitudes $T_{ij}$ in Eq.~\eqref{BS-eqn}, with $i,j=\Delta\pi, \Sigma^* K, N^*\pi, N\sigma$ and plot $|T_{ij}|^2$. We define the mass and width of the resonance the position of the peak and the width of the $|T_{ij}|^2$ distribution as a function of $\sqrt{s}$ close to the peak. In that region we have
\begin{equation}\label{eq:coupling1}
  T_{ij}=\frac{g_i \,g_j}{\sqrt{s}-M_R+i\frac{\Gamma_R}{2}}.
\end{equation}

We take the $\Sigma^* K$ channel as reference and then have
\begin{equation}\label{eq:g2}
\left.T_{22}\right|_{\rm peak}
  =\frac{g^2_2}{i\frac{\Gamma_R}{2}},~~~~~~~~~g^2_2=\left. i\frac{\Gamma_R}{2} \;T_{22}\right|_{\rm peak}.
\end{equation}
This defines $g_2$ up to an arbitrary sign, but then the rest of couplings are defined relative to this via
\begin{equation}\label{eq:gig2}
  \left.\frac{g_i}{g_2}=\frac{T_{i2}}{T_{22}}\right|_{\rm peak}.
\end{equation}

Once we have the coupling, the partial decay widths are given by
\begin{equation}\label{eq:Gammai}
  \Gamma_i=\frac{1}{2\pi}\frac{M_{B}}{M_R}\,|g_i|^2 p_i,
\end{equation}
where $M_B$ is the mass of the final baryon and $M_R$ the one of the resonance and
\begin{equation}\label{eq:pi}
  p_i=\frac{\lambda^{1/2}(M^2_R, M^2_m, M^2_B)}{2M_R}
\end{equation}
with $M_m$ the mass of the final meson in the channels $\Delta\pi, \Sigma^* K, N^*\pi, N\sigma$.

The width of all channels is well defined except for the $\Sigma^* K$ since the resonance is close to threshold and both theoretically and experimentally the determination of $\Sigma^* K$ in the width is uncertain. With this caveat, we shall check that the sum of all partial decay widths is close to the total width determined from the shape of $|T_{ij}|^2$.

\section{Results}\label{sec:results}
\begin{figure}[tbh]\centering
\includegraphics[scale=0.3]{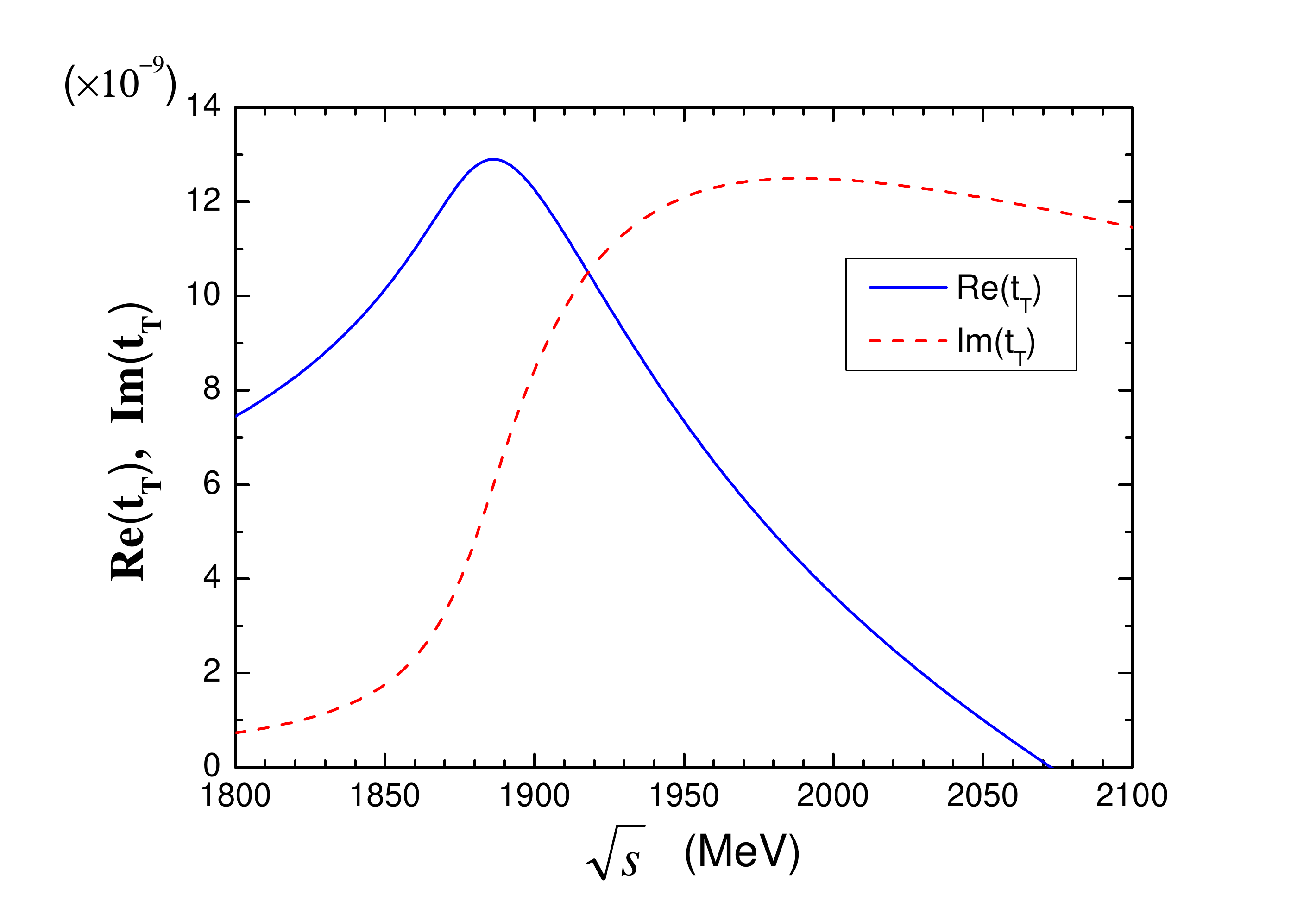}
\caption{Real and imaginary parts of $t_T$ of the triangle diagram, Eq. \eqref{t_T} with $\Sigma^* K$ intermediate state.
\label{fig:Fig8}}
\end{figure}

In Fig. \ref{fig:Fig8}, we show the results for $t_T$ as a function of $\sqrt{s}$ for $M_{N^*}\simeq 1535 \,{\rm MeV}$. We can see that ${\rm Re} (t_T)$ has a peak structure with a peak around $1885\,{\rm MeV}$. The  imaginary part has a different behaviour, and does not show any peak. Actually, $-it_T$ would resemble a Breit-Wigner amplitude with a constant magnitude added to the real part, which does not go through zero. The peak observed in ${\rm Re} (t_T)$ is tied to the triangle singularity that one would have in the case that $\Gamma_{\Sigma^*}\to 0$.

\begin{figure}[b!]\centering
\includegraphics[scale=0.28]{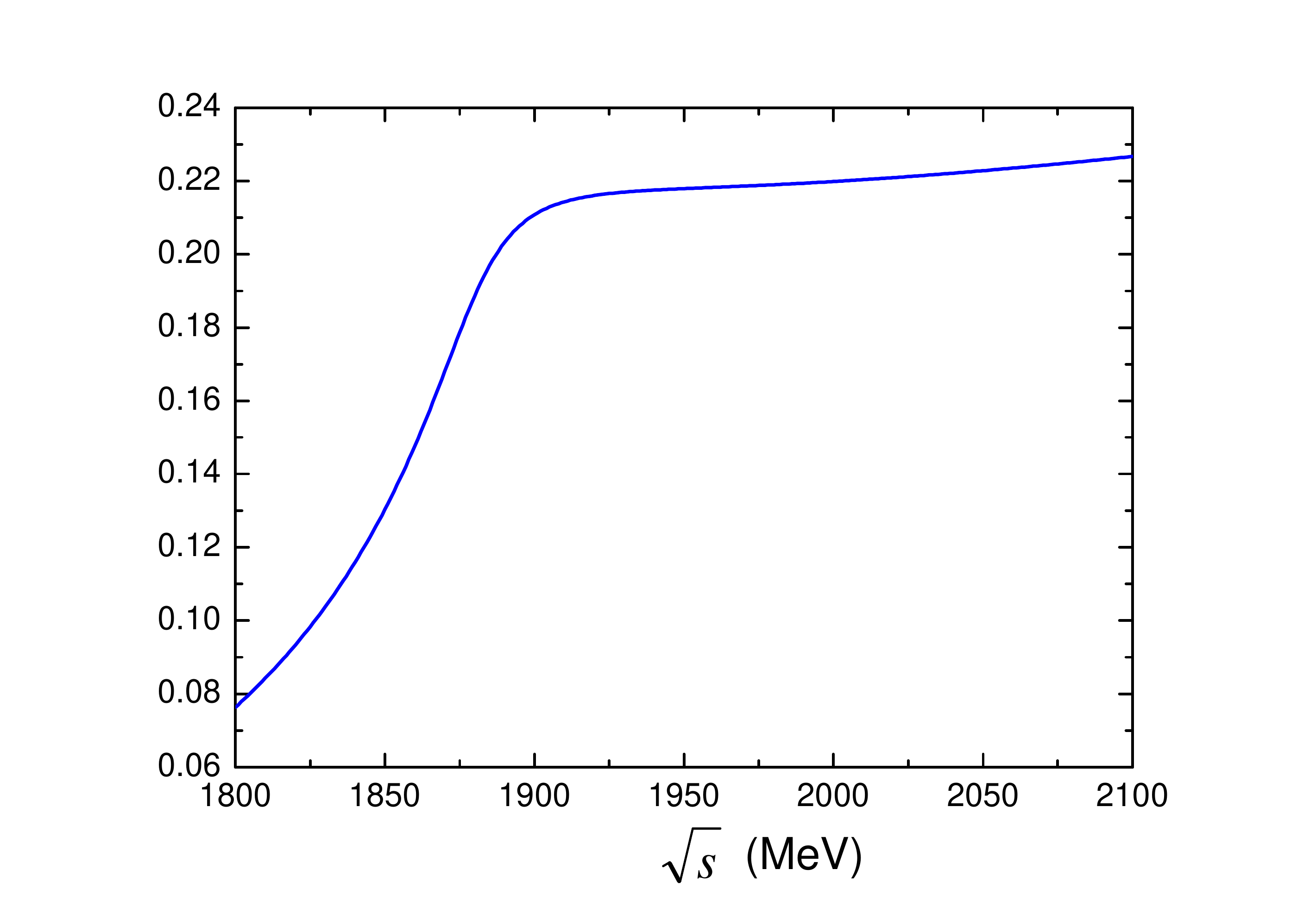}
\caption{$\tilde{V}_{i, \pi N^*}/V^{(I=1/2)}_{i, \Sigma^* K}$ of Eq. \eqref{eff-s-wave}.
\label{fig:Fig9}}
\end{figure}

In Fig. \ref{fig:Fig9} we show $\tilde{V}_{i, \pi N^*}/V^{(I=1/2)}_{i, \Sigma^* K}$ from Eq. (\ref{eff-s-wave}). This magnitude provides the relative strength of the effective transition potential $i\to \pi N^*$, with respect to $i\to \Sigma^* K$.
We observe that the effective potential rises rapidly up to $\sqrt{s}=1900\,{\rm MeV}$ and stabilizes there. The relative strength with respect to $V^{(I=1/2)}_{i, \Sigma^* K}$ is of the order of 0.22 at the peak, which anticipates a moderate effect of this channel. However, the added strength around $1880 \,{\rm MeV}$ helps stabilize the molecule that builds up around this energy from the interaction of the $\Delta \pi$ and $\Sigma^* K$ channels.

\begin{figure}[tbh]\centering
\includegraphics[scale=0.3]{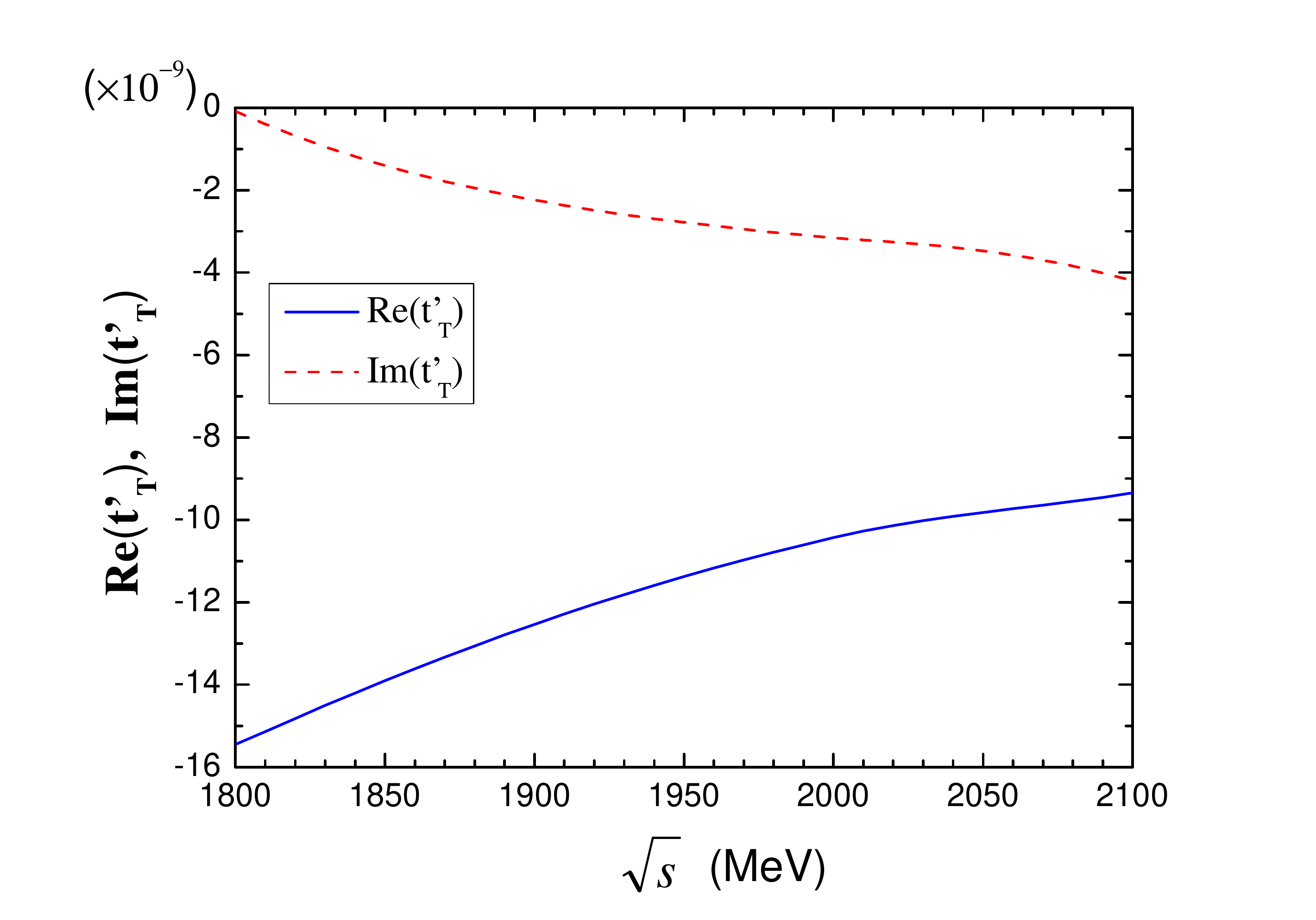}
\caption{Real and imaginary parts of $t'_T$ corresponding to the triangle diagram, of Fig. \ref{fig:Fig5}.
\label{fig:Fig10}}
\end{figure}

Next we show in Fig. \ref{fig:Fig10} the results for $t'_T$ of subsection \ref{subsec:Nf0_channel} for Fig. \ref{fig:Fig5}. The convolution of Eq. \eqref{eq:Imt'2} over the $\sigma$ mass is done between the masses $2m_\pi$ and $800\,{\rm MeV}$, and in Fig. \ref{fig:Fig10} we plot $t'_T$ in the middle of the range at $\tilde{m}_{\sigma}=540\,{\rm MeV}$. We can see that now we do not have any peak, as anticipated, since Eq. (\ref{TS-relation}), that shows when there is a triangle singularity, is not fulfilled in this case. Yet, we see that ${\rm Re} (t'_T)$ is of the same order of magnitude as ${\rm Re}(t_T)$ at the peak. However, since the effective transition potential contains different couplings now, its strength becomes bigger than the one of the $\Sigma^* K$ in the loop, as we show below.

\begin{figure}[h!]\centering
\includegraphics[scale=0.25]{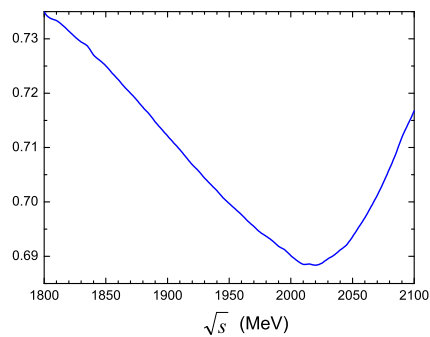}
\caption{$\tilde{V}_{i, \sigma N}/V^{(I=1/2)}_{i, \Delta \pi}$ as a function of the energy corresponding to Eq. \eqref{eq:V_effective4}.
\label{fig:Fig11}}
\end{figure}

In Fig. \ref{fig:Fig11} we plot $\tilde{V}_{i, \sigma N}/V^{(I=1/2)}_{i, \Delta \pi}$.
We can see that this magnitude is relatively constant, and from $1800\,{\rm MeV}$ to $2100\,{\rm MeV}$ it changes from 0.73 to 0.69. However, we can see now that the strength is bigger than the one obtained from the $\Sigma^* K$ loop at its peak ($\sim 0.22$), in spite of the fact that we do not have a singularity now. As mentioned before, the different couplings in the mechanism are responsible for this relatively large strength. We see that the strength of $\tilde{V}_{i, \sigma N}$ is of the same order of magnitude as the $V^{(I=1/2)}_{i, \Delta \pi}$ transitions and one anticipates an important role of this channel.

Next we turn to the amplitudes obtained with the coupled channels problem.

\begin{figure}[bp!]\centering
\includegraphics[scale=0.325]{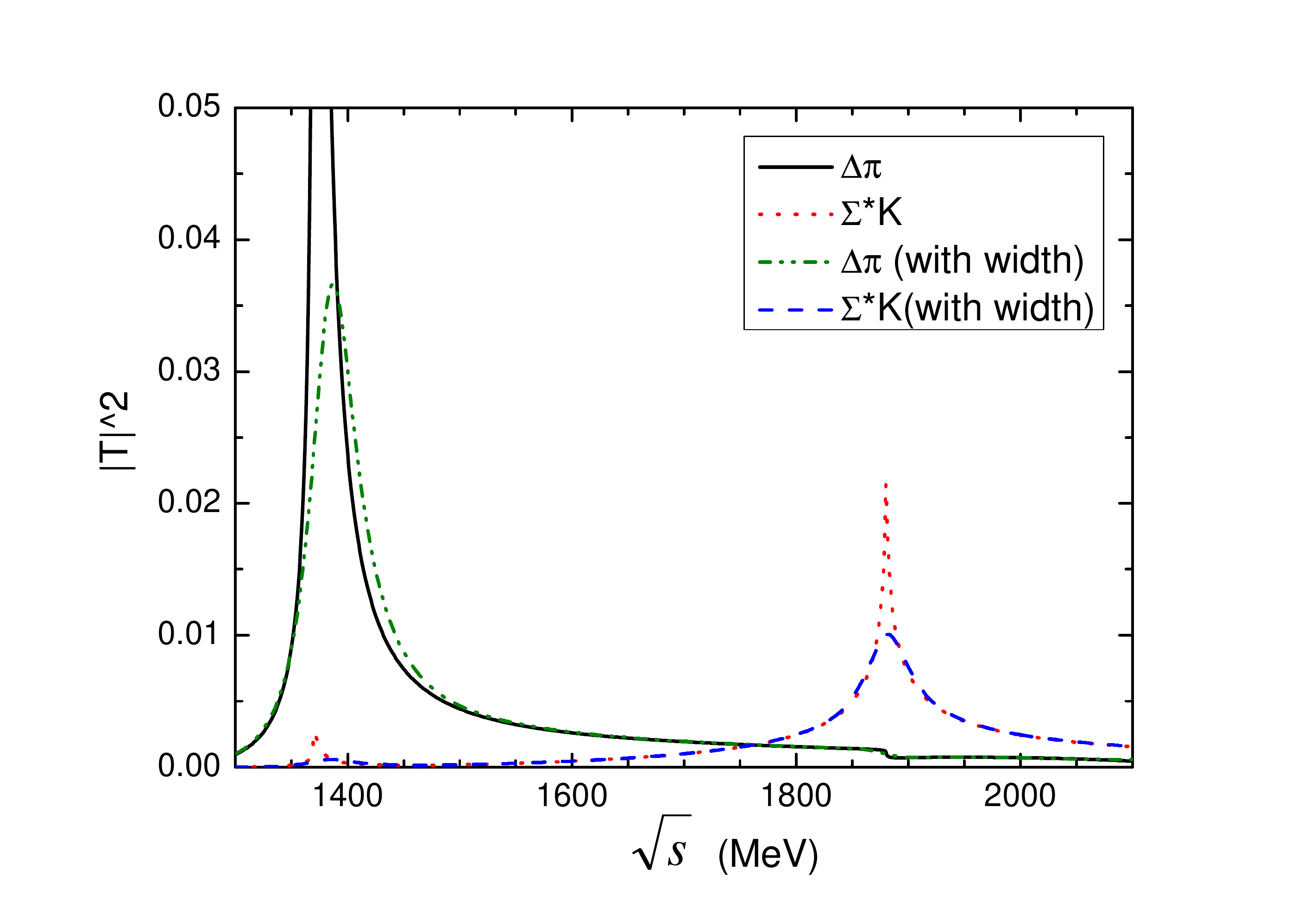}
\caption{$|T_{ii}|^2$ for the $\Delta \pi$ and $\Sigma^* K$ channels alone as a function of the energy. The wider curve corresponds to considering the width of $\Delta$ and $\Sigma^*$.
\label{fig:Fig12}}
\end{figure}

In Fig. \ref{fig:Fig12} we show the module square of amplitude $T_{ii}$ (with the order of the channels, $\Delta\pi, \Sigma^* K, N^*\pi, N\sigma$) with just the $\Delta\pi$ and $\Sigma^* K$ channels omitting the width of the $\Delta$ and $\Sigma^*$, and taking it into account.
 The results are similar to those obtained in Ref.~\cite{sarkar}, though in Ref.~\cite{sarkar} complex energies were used instead of the convolution in the evaluation of the $G$ function of Eq. (\ref{BS-eqn}). We can see a clear peak around $1880\,{\rm MeV}$ and that the consideration of the width of the $\Delta$ and $\Sigma^*$ leads to a wider structure which has about $72\,{\rm MeV}$, short of the experimental central value of about $200\,{\rm MeV}$, which, however, has large uncertainties.

\begin{figure}[bp!]\centering
\includegraphics[scale=0.325]{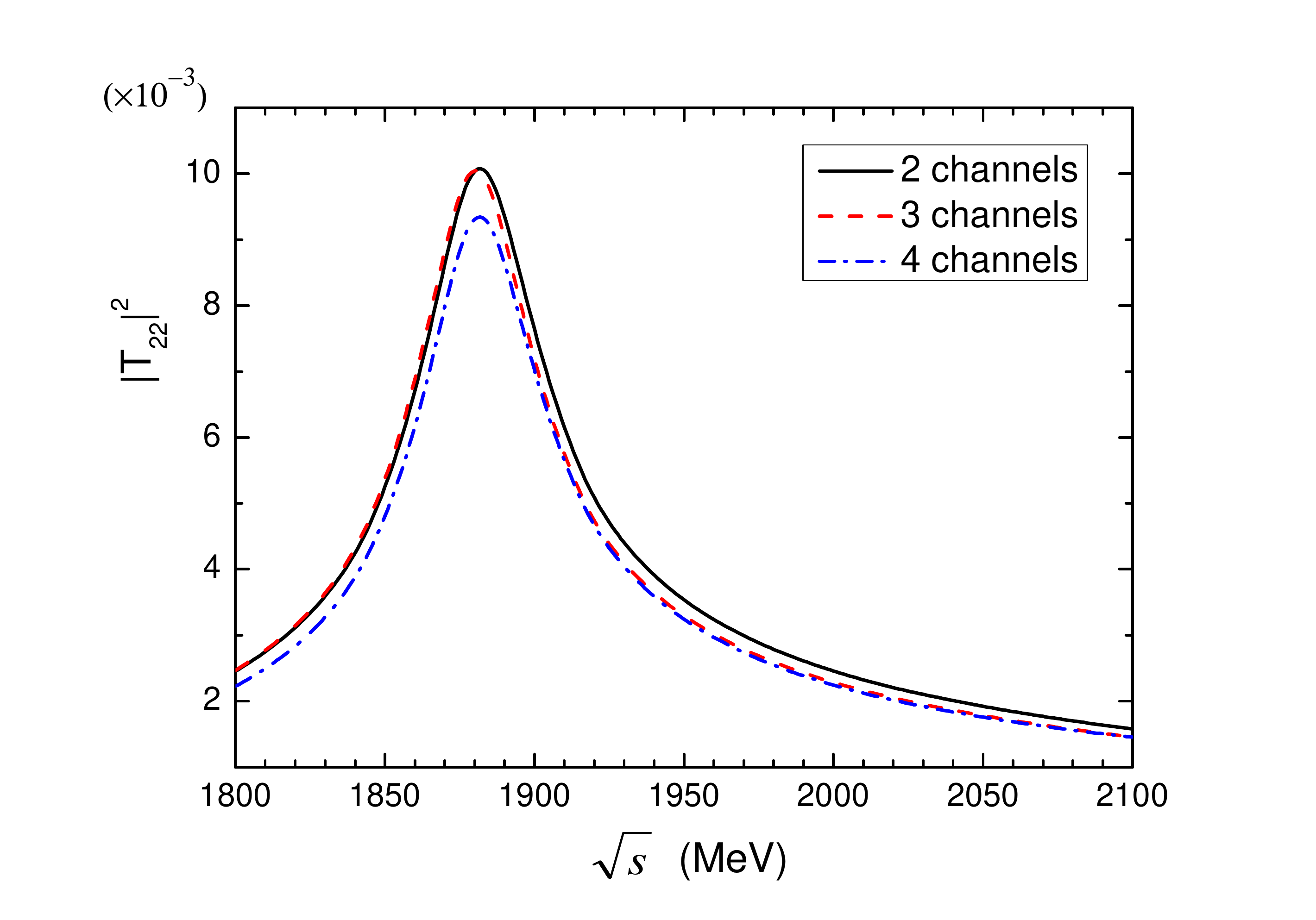}
\caption{$|T_{22}|^2$ with two channels ($\Delta\pi, \Sigma^* K$), three channels ($\Delta\pi, \Sigma^* K, N^* \pi$) and four channels ($\Delta\pi, \Sigma^* K, N^* \pi, N\sigma$).
\label{fig:Fig13}}
\end{figure}

In Fig. \ref{fig:Fig13} we show again the module square of $T_{22}$ amplitude with two channels ($\Delta\pi, \Sigma^* K$), three channels ($\Delta\pi, \Sigma^* K, N^* \pi$) and four channels ($\Delta\pi, \Sigma^* K, N^* \pi, N\sigma$).
We can see that the introduction of the $N^* \pi$ channel widens the peak a bit.
The introduction of the $N\sigma$ channel has also not much effect on the width, but we shall see later that it has an important repercussion in the $\pi\pi$ invariant mass distribution.
From $|T_{22}|^2$ with four channels, we can get the mass and width of the $N^*(1875)$ resonance:~ $M_R=1881.7\;{\rm MeV}$,  $\Gamma_R=71.2\;{\rm MeV}$.

\begin{figure}[tbp]\centering
\includegraphics[scale=0.32]{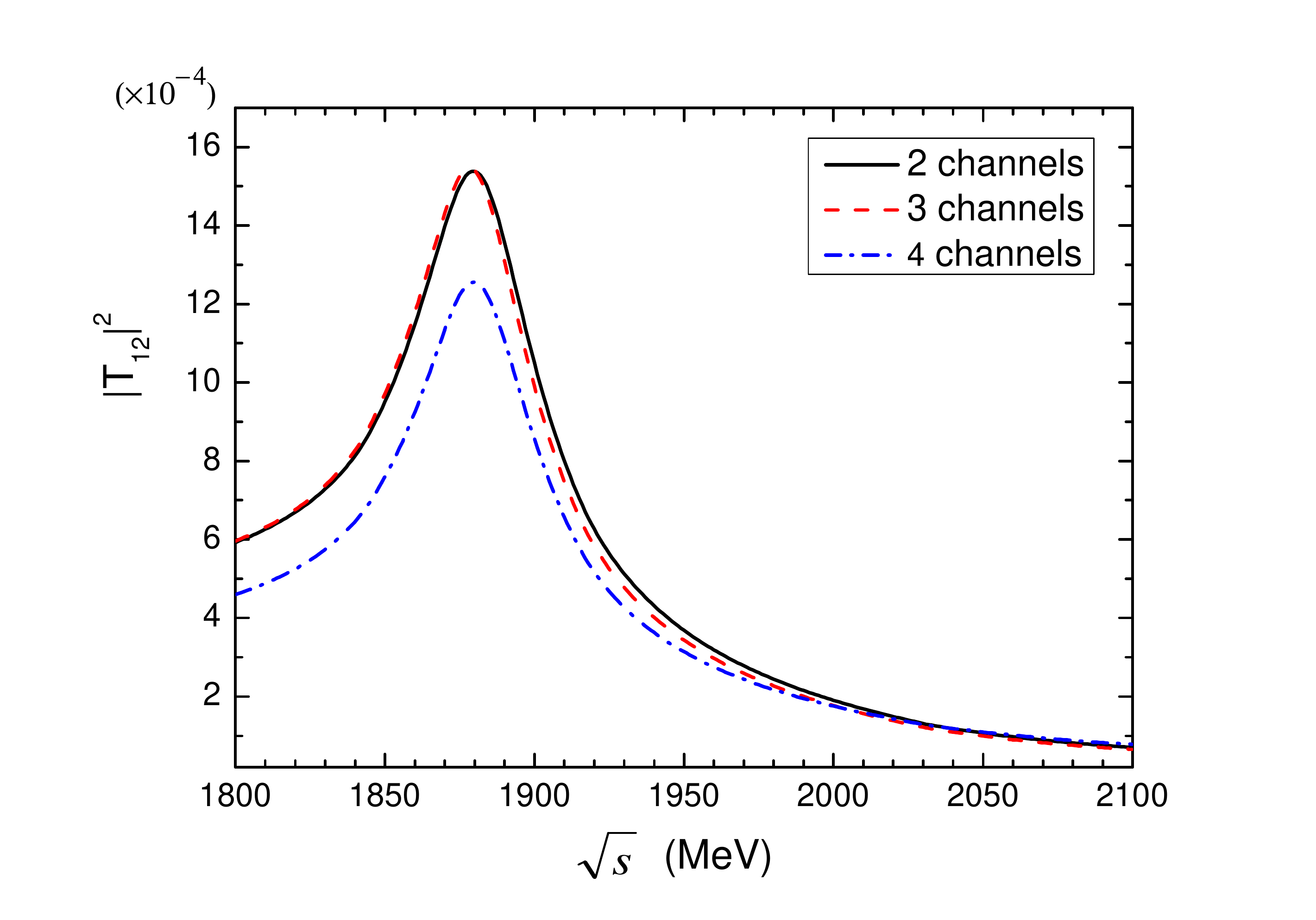}
\caption{$|T_{12}|^2$ as a function of the energy, for $\Delta \pi \to \Sigma^* K$ transition.
\label{fig:Fig14}}
\end{figure}

\begin{figure}[tbp]\centering
\includegraphics[scale=0.32]{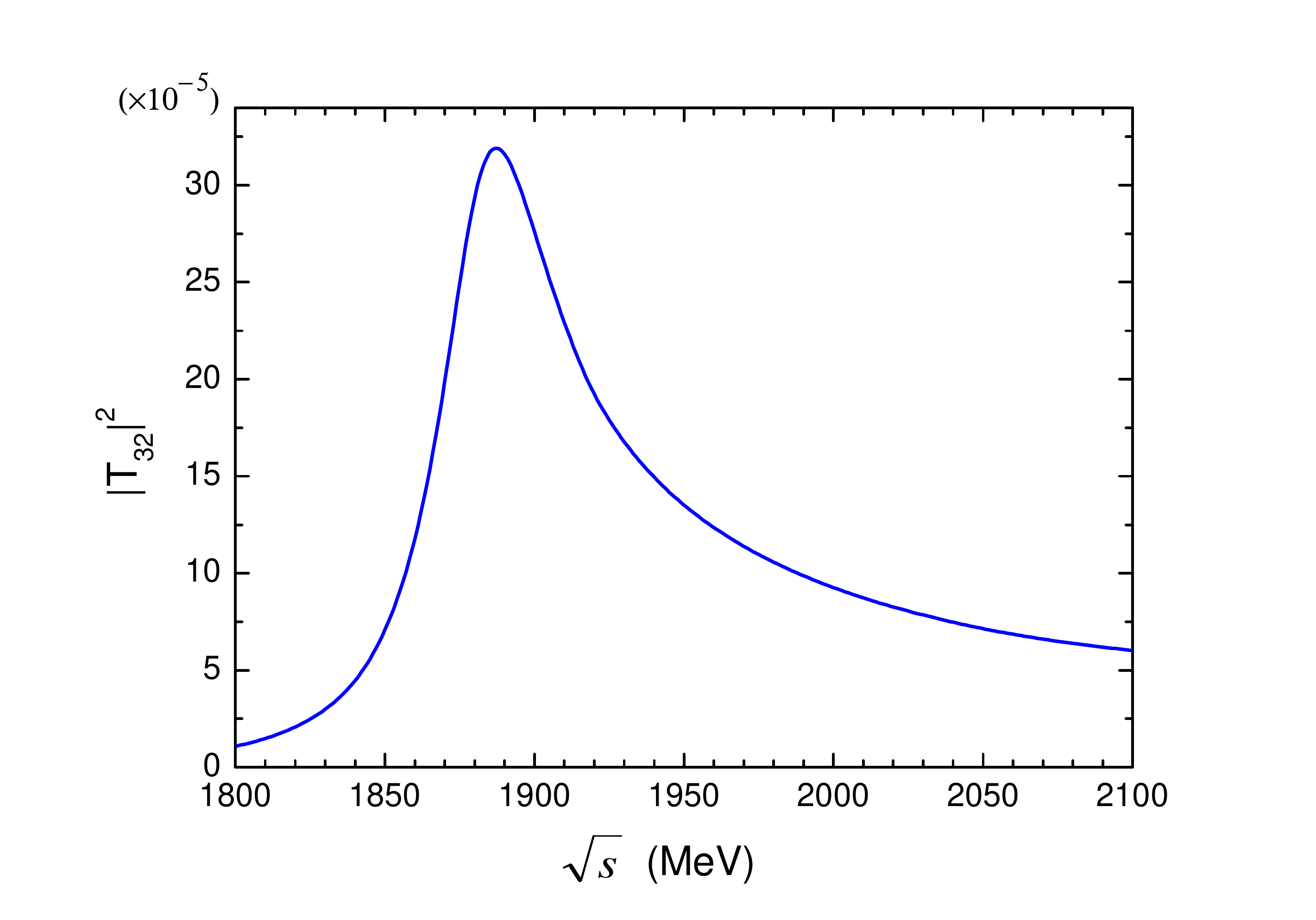}
\caption{$|T_{32}|^2$ as a function of the energy for $\Sigma^* K \to N^* \pi$ transition.
\label{fig:Fig15}}
\end{figure}

\begin{figure}[tbp]\centering
\includegraphics[scale=0.32]{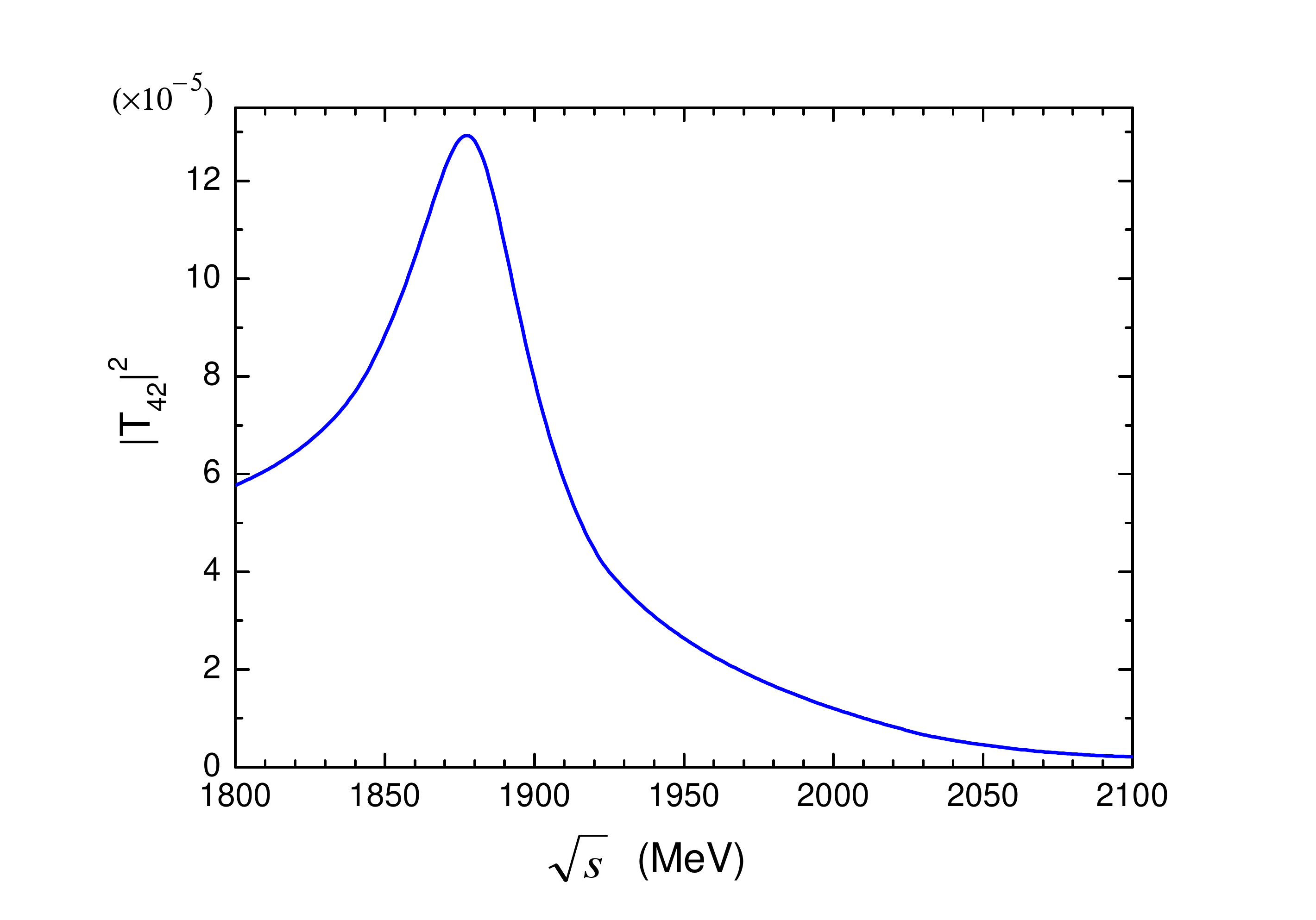}
\caption{$|T_{42}|^2$ as a function of the energy for $\Sigma^* K \to N\sigma$ transition.
\label{fig:Fig16}}
\end{figure}

Next we look at the transition amplitudes from where we determine the couplings, via Eqs. \eqref{eq:g2} and \eqref{eq:gig2}. We show $|T^2_{12}|$ in Fig. \ref{fig:Fig14}, $|T^2_{32}|$ in Fig. \ref{fig:Fig15} and $|T^2_{42}|$ in Fig. \ref{fig:Fig16}, all of them evaluated with the four channels.

The couplings that we get using Eqs. \eqref{eq:g2} and \eqref{eq:gig2} are:
\begin{align}\label{eq:couplings}
   & g_{R, \Sigma^* K}=1.72 - 0.70i,~~~ &  g_{R, \Delta \pi}=0.34 + 0.59i, \nonumber\\
   & g_{R, N^* \pi}=-0.29 + 0.17i,    &  g_{R, N \sigma}=0.22 - 0.01i.
\end{align}
With these values
and using $\Gamma_i = \frac{1}{2\pi}\frac{M_i}{M_R}|g_i|^2p_i$, where $M_i$ is the baryon mass for the final channel $i$ and $p_i$ its momentum,
 we obtain the partial decay widths
\begin{align}\label{eq:widths}
   & \Gamma_{\Delta \pi}=25.2~{\rm MeV}, &  \Gamma_{\Sigma^* K}=13.1~{\rm MeV}, \nonumber\\
   & \Gamma_{N^* \pi}=4.2~{\rm MeV},~~~~~    &  \Gamma_{N \sigma}=2.3~{\rm MeV}.
\end{align}
We can see that $\Gamma_{\Delta \pi}$ is quite large,   but $\Gamma_{N^* \pi}$ and $\Gamma_{N \sigma}$ are much smaller.

The sum of $\Gamma_i$ is $44.8~{\rm MeV}$, much smaller than the total width $\Gamma_R=71.2\;{\rm MeV}$.
Yet, since  the peak of the $N^*(1875)$ has a mass distribution and the $\Sigma^*$ has a width $\Gamma_{\Sigma^*}=36~{\rm MeV}$ \cite{pdg}, we should do a double convolution for the partial decay width $\Gamma_{\Sigma^* K}$,
\begin{eqnarray}\label{eq:doublecon}
  \tilde{\Gamma}_{\Sigma^* K} \!\!&=&\!\! \frac{1}{N_R}\int^{M_R+2\Gamma_R}_{M_R-2\Gamma_R} {\rm d}\tilde{M}_R\; \int^{M_{\Sigma^*}+2\Gamma_{\Sigma^*}}_{M_{\Sigma^*}-2\Gamma_{\Sigma^*}}{\rm d}\tilde{M}_{\Sigma^*} \nonumber\\
   \!\!&\times&\!\! S_R(\tilde{M}_R) \;  S_{\Sigma^*}(\tilde{M}_{\Sigma^*}) \; \Gamma_{\Sigma^* K}(\tilde{M}_R, \tilde{M}_{\Sigma^*}, m_K),~~~
\end{eqnarray}
where $S_R(\tilde{M}_R)$ (or $S_{\Sigma^*}(\tilde{M}_{\Sigma^*})$) is the spectral function of $N^*(1875)$ (or $\Sigma^*$), taking the same form of Eq.~\eqref{eq:specFun} with proper mass and width for the resonance; and
\begin{equation*}
  N_R\!=\!\!\int^{M_R+2\Gamma_R}_{M_R-2\Gamma_R}\!\! {\rm d}\tilde{M}_R \!\!\int^{M_{\Sigma^*}+2\Gamma_{\Sigma^*}}_{M_{\Sigma^*}-2\Gamma_{\Sigma^*}}\!\!{\rm d}\tilde{M}_{\Sigma^*} S_R(\tilde{M}_R) \,  S_{\Sigma^*}(\tilde{M}_{\Sigma^*}),
\end{equation*}
\begin{equation}\label{eq:GamSigK}
  \Gamma_{\Sigma^* K}(\tilde{M}_R, \tilde{M}_{\Sigma^*}, m_K)=\frac{1}{2\pi} \frac{\tilde{M}_{\Sigma^*}}{\tilde{M}_R} \; g^2_{R, \Sigma^* K} \;\tilde{p},
\end{equation}
with
\begin{equation*}
  \tilde{p}=\frac{\lambda^{1/2}(\tilde{M}^2_R, \tilde{M}^2_{\Sigma^*}, m^2_K)}{2\tilde{M}_R} \; \theta(\tilde{M}_R - \tilde{M}_{\Sigma^*}-m_K ).
\end{equation*}
We then get
\begin{equation}\label{eq:finGamSigK}
\tilde{\Gamma}_{\Sigma^* K}= 33.2~{\rm MeV}.
\end{equation}
Then the sum of partial decay widths is $64.9~{\rm MeV}$, compatible with the total width.

For $\Gamma_{N^* \pi}$ the prediction is new and should be observed in the $\pi \eta N$ mode since $N^*(1535)$ decays into $\eta N$ with a branching fraction of 32-52\% and it would be a better channel than the $\pi N$ that could be mixed with the $\Delta \pi$ decay. As to $\Sigma^* K$, which is also not measured, there are certainly problems when one is close to threshold. However, a proper unitary multichannel analysis, as done in
Ref.~\cite{manley,misha,sato}, should show the relevance of this channel. One similar case where this has been done is in the $N^*(1700)(3/2^-)$ resonance, which in Ref.~\cite{angelsvec} is shown to appear from the interaction of vector-baryon, mostly from $\rho N$, which is at threshold there. This case has been revised in Ref.~\cite{singuroca} to include the $\Delta \pi$ channel, associated to another triangle singularity. The $\rho N$ channel being around threshold is not an obstacle to obtain a $(38 \pm 6)$\% branching ratio for $\rho N$ in the analysis of Ref.~\cite{manley}.

\begin{figure*}[htbp]\centering
\includegraphics[scale=0.53]{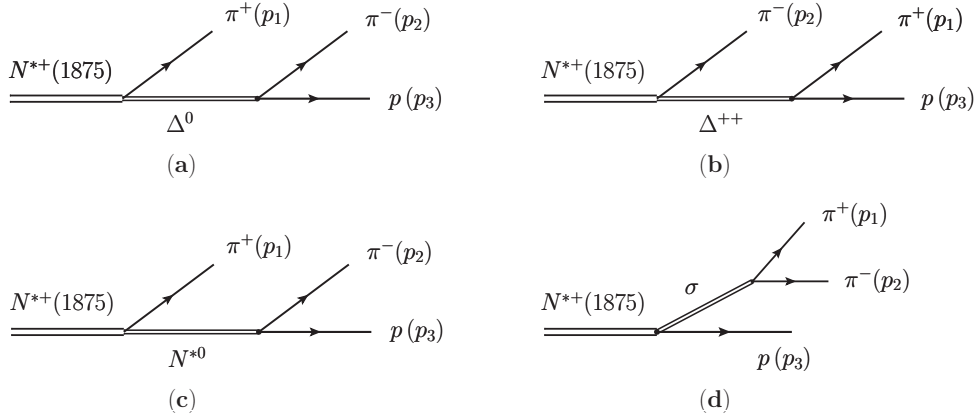}
\caption{Diagrams for the decay of the $N^*(1875)$ resonance into $\pi \Delta$ [(a) and (b)], $\pi N^*$ [(c)] and $\sigma p$ [(d)].
\label{fig:Fig17}}
\end{figure*}

\section{Mass distributions}
Now we wish to get the mass distributions for pairs of particles. We choose to study the $\pi^+\pi^-p$ final state. We will have the contributions of Fig. \ref{fig:Fig17}.
The first thing to observe is that the $\pi^+\pi^-p$ channel does not exhaust all the width. Indeed, in the case of $\pi\Delta$ decay we have three more cases $R^+\rightarrow \pi^+\Delta^0\rightarrow \pi^+\pi^0n$, $R^+\rightarrow \pi^0\Delta^+\rightarrow \pi^0\pi^+n$ and $R^+\rightarrow \pi^0\Delta^+\rightarrow \pi^0\pi^0p$, with $R$ standing for the resonance $N^*(1875)$. Using the coefficients for the weights of the different $\pi\Delta$ components in Eq. (\ref{isospindecom}) and those for $\Delta\rightarrow \pi N$ in Eq. (\ref{eq:Ci}) we find that the $\pi^+\pi^-p$ mechanisms of Fig. \ref{fig:Fig17} account for $\frac59$ of the $\pi\Delta$ width, while the channels not considered account for $\frac49$ of the width. Similarly, for the $\pi N^*$ we are missing $R^+\rightarrow \pi^+ N^{*0}\rightarrow \pi^+\pi^0n$, $R^+\rightarrow \pi^0 N^{*+}\rightarrow \pi^0\pi^+n$ and $R^+\rightarrow \pi^0 N^{*+}\rightarrow \pi^0\pi^0p$. Taking into account the coefficients of Eq. (\ref{isospindecom}), we find again that with the $\pi^+\pi^-p$ final state we take into account $\frac49$ of the $\pi N^*$ width and the missing channels account for $\frac59$ of it. As to the spin dependence of the diagrams of the Fig. \ref{fig:Fig17}, the $R^+(\frac32^-)$ coupling to $\pi\Delta$ goes as a constant, and the $\Delta\rightarrow \pi N$ as $\vec S\cdot\vec p_\pi$. The $N^*\rightarrow \pi N$ also goes as a constant but the $R\rightarrow \pi N^*$ goes as $\vec S\cdot\vec p_\pi$. On the other hand the $R^+(\frac32^-)$ coupling to $\sigma p$ also goes as $\vec S\cdot\vec p_\pi$ (see Eq. (\ref{eq:Ci})). Then considering Eqs. (\ref{isospindecom}) and (\ref{eq:Ci}) and the isospin decomposition of the $N^{*0}\rightarrow \pi N$ decay, $|N^{*0}, \pi N, I=\frac12, I_3=-\frac12\rangle = \sqrt{\frac13}|\pi^0 n\rangle - \sqrt{\frac23}|\pi^-p\rangle$, we have
\begin{eqnarray}\label{mass-distribut-amp}
-it_{R^+ \to \pi^+\Delta^0\to \pi^+\pi^-p}\!\! &=&\!\! g_{R,\pi\Delta}(-\sqrt{\frac16})\sqrt{\frac13}
\nonumber\\
\!\!&\times&\!\! \!\frac{1}{M_{23}- M_\Delta + i\frac{\Gamma_\Delta}{2}}\vec S\cdot\vec p_2\frac{f_{\pi N\Delta}}{m_\pi},
\nonumber\\
-it_{R^+ \to\pi^-\Delta^{++}\rightarrow \pi^-\pi^+p}\!\! &=&\!\! g_{R,\pi\Delta}\sqrt{\frac12}(-1)\vec S\cdot\vec p_1
\nonumber\\
\!\!&\times&\!\!\! \frac{1}{M_{13}- M_\Delta + i\frac{\Gamma_\Delta}{2}}\frac{f_{\pi N\Delta}}{m_\pi},\\
-it_{R^+ \to\pi^+N^{*0}\rightarrow \pi^+\pi^-p}\!\! &=&\!\! g'_{R,\pi\Delta}(\sqrt{\frac23})(-\sqrt{\frac23})\vec S\cdot\vec p_1
\nonumber\\
\!\!&\times&\!\!\! \frac{1}{M_{23}- M_{N^*} + i\frac{\Gamma_{N^*}}{2}} g_{N^*,\pi N},
\nonumber\\
-it_{R^+ \to\sigma p\rightarrow \pi^+\pi^-p}\!\! &=& \!\!g'_{R,\sigma N}(-\sqrt{\frac23})\vec S\cdot(\vec p_1 + \vec p_2 )(-1)
\nonumber\\
\!\!&\times&\!\!\! \frac{1}{M_{12}^2- m_{\sigma}^2 + i m_{\sigma}\Gamma_{N^*}} g_{\sigma,\pi \pi},\nonumber
\end{eqnarray}
where $M_{ij}^2 = (p_i + p_j)^2$; $i,j=1,2,3$ for $\pi^+,\,\pi^-,\,p$.
In the last equation we have considered that $|\pi\pi, I=0\rangle = -\frac{1}{\sqrt{3}}\big(\pi^+\pi^- + \pi^-\pi^+ + \pi^0\pi^0\big)$. The coupling of the $\sigma$ to $\pi^+ \pi^-$ has a $(-\frac{1}{\sqrt{3}})$ coefficient, and so for the $\pi^-\pi^+$, but considering the integrated $\pi^+\pi^-$ and $\pi^-\pi^+$ width one is counting twice the contribution. All this is solved by taking the coefficient $(-\sqrt{\frac23})$. The extra minus sign in the last of Eqs. (\ref{mass-distribut-amp}) is because $p_p = -\vec p_1 - \vec p_2$.

We take $g_{N^*,\pi N}=0.70$ from Ref.~\cite{inoue}. In Eq. (\ref{mass-distribut-amp}), we have used the coupling $g'_{R,\pi N^*}$ and $g'_{R,\sigma N}$ instead of $g_{R,\pi N^*}$ and $g_{R,\sigma N}$. This is because the factors $\vec S\cdot\vec k$ were already taken into account when we evaluated the effective transition potentials (see Eq. (\ref{s-d wave}) and Eqs. (\ref{Imttildeij}) and (\ref{eq:Imt'2})), which already incorporate the factor $\frac13\vec k^{\,2}$ coming from this operator in the sum over $\pi N^*$ and $\sigma p$ intermediate states. To take this into account it is sufficient to write
\begin{equation}\label{eq:g'R}
  g'_{R,\pi N^*} = \frac{\sqrt{3}}{p_1}g_{R,\pi N^*},~~~~~g'_{R,\sigma p} = \frac{\sqrt{3}}{|\vec p_1 + \vec p_2|}g_{R,\sigma p}.
\end{equation}
After this discussion, we can write the full amplitude for $R^+\rightarrow \pi^+\pi^- p$ from the diagrams of Fig. \ref{fig:Fig17} as
\begin{eqnarray}\label{eq:t_tot}
-it_{\rm tot} &=& (B + C + D)\vec S\cdot \vec p_1 + (A + D)\vec S\cdot \vec p_2
\nonumber\\
&=& A' \;\vec S\cdot \vec p_1 + B'\; \vec S\cdot \vec p_2,
\end{eqnarray}
where
\begin{eqnarray}\label{def-ABCD}
A\!\! &=&\!\! -\frac{1}{3\sqrt{2}}\frac{g_{R,\pi\Delta} \cdot f_{\pi N \Delta}}{m_\pi}\frac{1}{M_{23} - M_\Delta + i\frac{\Gamma_\Delta}{2}},
\nonumber\\
B \!\!&=&\!\! -\frac{1}{\sqrt{2}}\frac{g_{R,\pi\Delta}\cdot f_{\pi N \Delta}}{m_\pi}\frac{1}{M_{13} - M_\Delta + i\frac{\Gamma_\Delta}{2}},\nonumber\\
C \!\!&=&\!\! -\frac{2}{\sqrt{3}}\frac{g_{R,\pi N^*}\cdot g_{N^*,\pi N}}{p_1}\frac{1}{M_{23} - M_{N^*} + i\frac{\Gamma_{N^*}}{2}},
\\
D \!\!&=&\!\! \sqrt{2}\;\frac{g_{R,\sigma N}\cdot g_{\sigma,\pi\pi}}{|\vec p_1 + \vec p_2|}\frac{1}{M^2_{12} - m^2_{\sigma} + im_\sigma \Gamma_{\sigma} },
\nonumber\\
A' \!\!&=&\!\! B+C+D, \quad\quad\quad   B' = A+D. \nonumber
\end{eqnarray}
The differential mass distribution is give by \cite{pdg}
\begin{eqnarray}\label{diff-Gamma}
\frac{{\rm d}^2 \Gamma}{{\rm d} M_{12} {\rm d} M_{23}} \!=\! \frac{1}{(2\pi)^3}\frac{4M_p M_R}{32M_R^3}\overline{\sum}\sum \Big| t_{\rm tot}\Big|^2 4M_{12} M_{23},~~
\end{eqnarray}
where using Eq. (\ref{s-d wave}), we find
\begin{eqnarray}
&&\overline{\sum}\sum \Big| t_{\rm tot}\Big|^2 \nonumber\\
&=& \frac13\Big[ |A'|^2 \vec p_1^{\,2} + |B'|^2\vec p_2^{\,2}
+2 {\rm Re}(A'B'^*)\vec p_1\cdot\vec p_2 \Big],
\end{eqnarray}
where $\vec p_1\cdot \vec p_2$ can be written in terms of $M_{12}$ as
\begin{eqnarray}
2\vec p_1\cdot\vec p_2 = m_{1}^2 + m_{2}^2 + 2E_1E_2 - M_{12}^2
\end{eqnarray}
and $E_1$, $E_2$ as
\begin{eqnarray}
E_1 = \frac{M_R^2 + m_1^2 - M_{23}^2}{2M_R},~~~ E_2 = \frac{M_R^2 + m_2^2 - M_{13}^2}{2M_R}.
\end{eqnarray}
To obtain $\frac{{\rm d}\Gamma}{{\rm d}M_{12}}$, we integrate Eq. (\ref{diff-Gamma}) over $M_{23}$ and the limits are given in the PDG \cite{pdg}. In $t_{\rm tot}$ we need $M_{12}$, $M_{13}$, $M_{23}$ as variables. To evaluate Eq. (\ref{diff-Gamma}), we need $M_{13}$ which is given in terms of the other variables as
\begin{eqnarray}
M_{13}^2 = M_R^2 + m_1^2 + m_2^2 + m_3^2 - M_{12}^2 - M_{23}^2.
\end{eqnarray}
If we wish to obtain $\frac{{\rm d}\Gamma}{{\rm d}M_{23}}$. We integrate Eq. (\ref{diff-Gamma}) over $M_{12}$. The limits for $M_{23}$ can be obtained from those for $M_{12}$ by permuting the indices $123\rightarrow 321$. Similarly we can obtain $\frac{{\rm d}^2\Gamma}{{\rm d}M_{12}{\rm d}M_{13}}$ as in Eq. (\ref{diff-Gamma}) substituting the factor $2M_{23}$ by $2M_{13}$. Then we get $\frac{{\rm d}\Gamma}{{\rm d}M_{13}}$ integrating over $M_{12}$ and and limits for $M_{12}$ are obtained from the standard formula of the PDG permuting the indices $123\rightarrow 312$.

\section{Results for the mass distributions}
In the limit of the small widths for the $\Delta$, $N^*$ and $\sigma$, the different terms in Eq. (\ref{mass-distribut-amp}) do not interfere since they correspond to different final states $\pi\Delta$, $\pi N^*$, $\sigma N$. However, if we look at $\pi^+\pi^- p$ production and consider the widths there can be interference. In particular there should be interference between $\pi^-\Delta^{++}$ and $\sigma p$ ($B$ and $D$ terms in Eqs. (\ref{def-ABCD})). Note that the $B$ term is three times larger in strength than term $A$). The fact that these two terms have the same spin structure $(\vec S\cdot\vec p_1)$ helps for the interference.

\begin{figure}\centering
\includegraphics[scale=0.35]{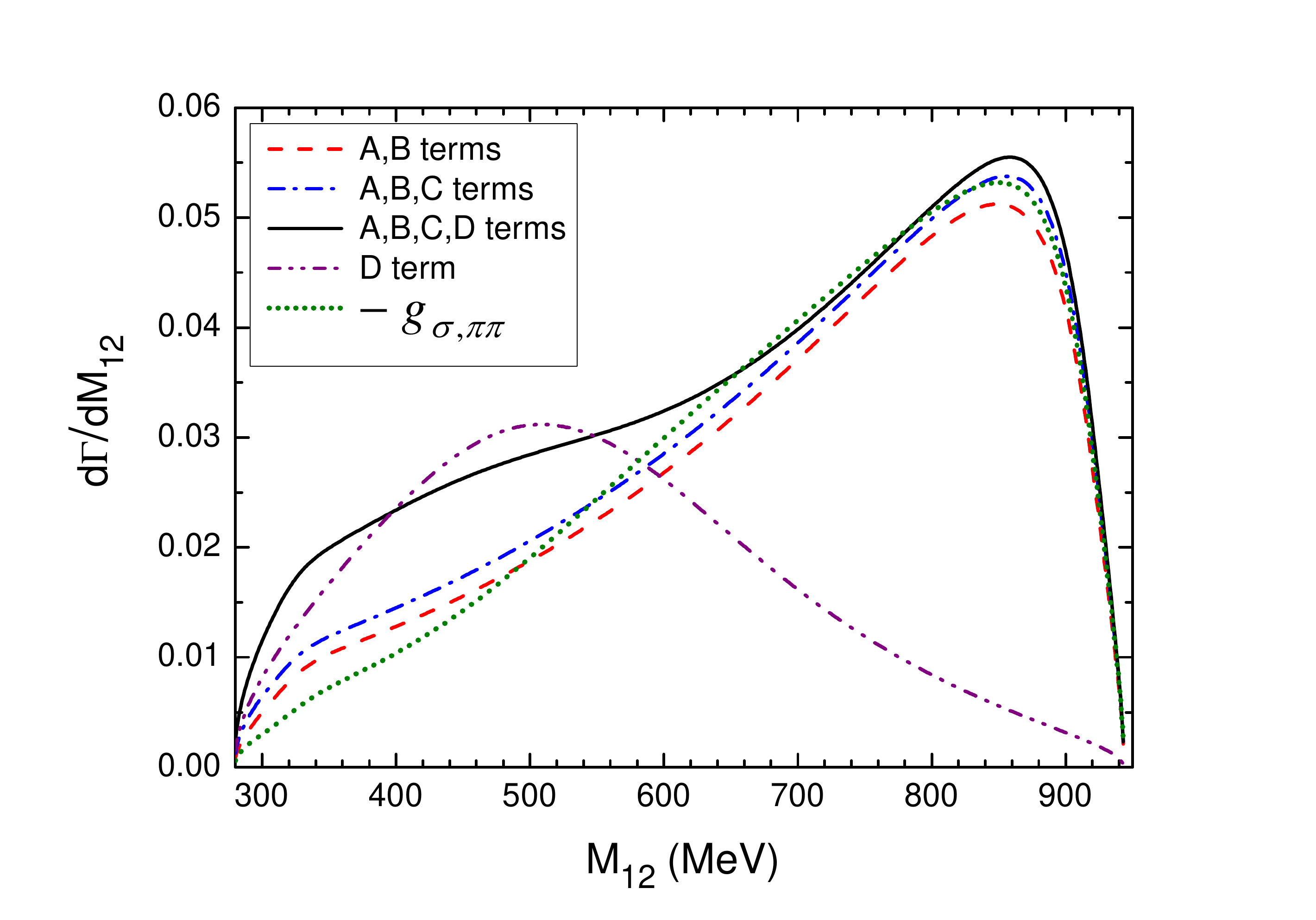}
\includegraphics[scale=0.35]{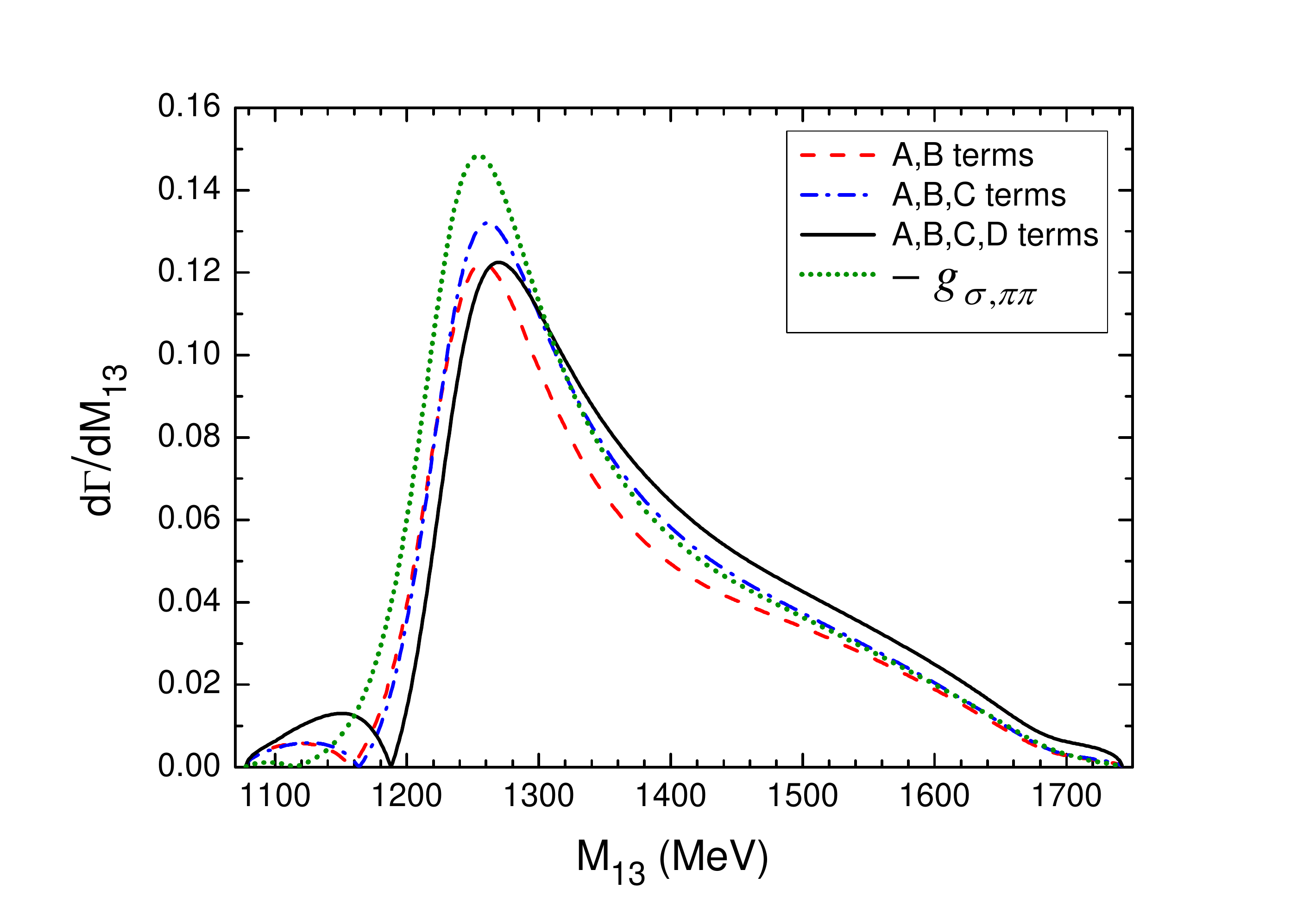}
\includegraphics[scale=0.35]{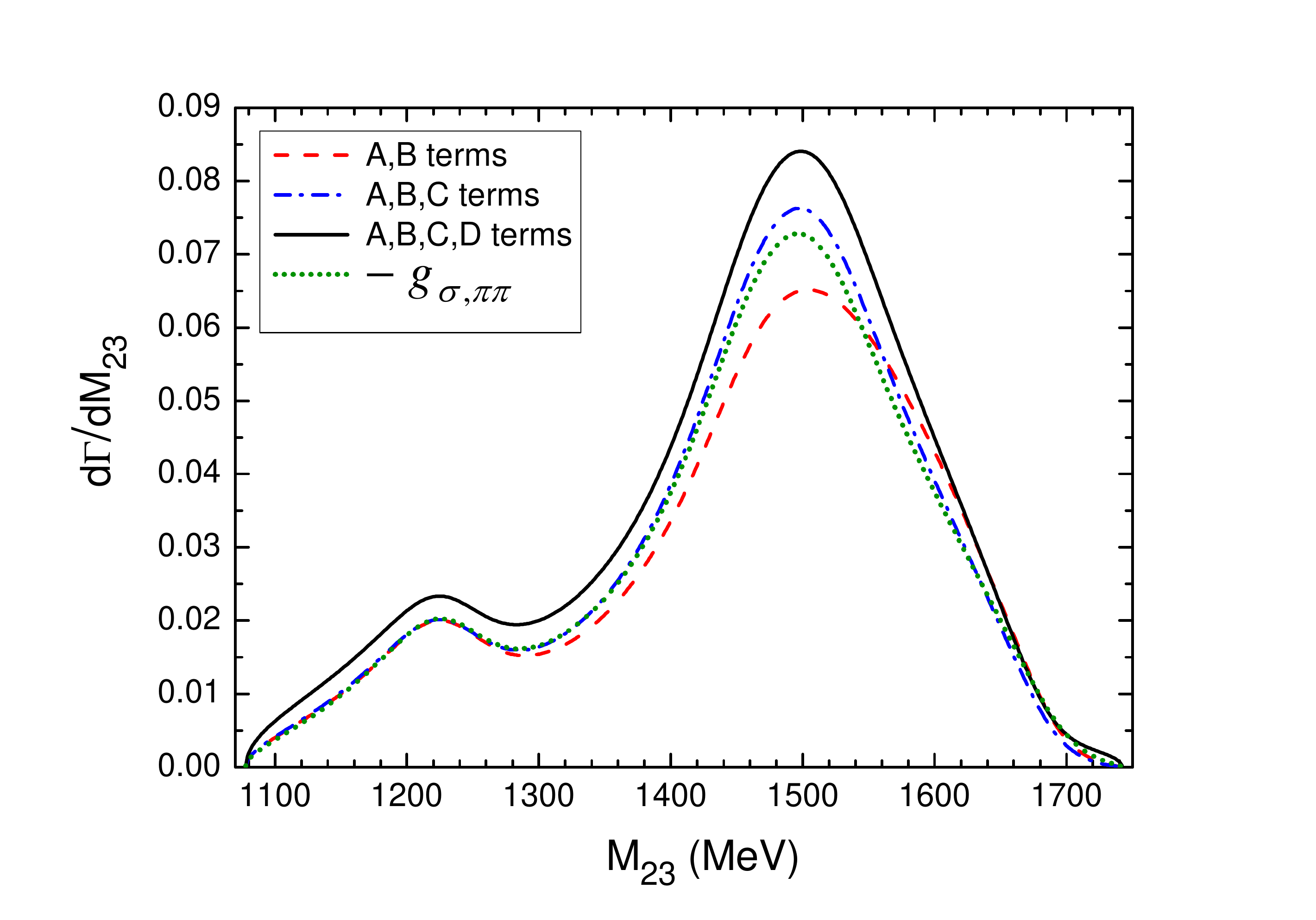}
\caption{The $\frac{{\rm d}\Gamma}{{\rm d}M_{12}}$, $\frac{{\rm d}\Gamma}{{\rm d}M_{13}}$ and $\frac{{\rm d}\Gamma}{{\rm d}M_{23}}$ mass distributions for the $N^{*+}(1875) \to \pi^+\pi^- p$ decay with 1,2,3 denoting $\pi^+, \pi^-$ and $p$,
 with the red dashed lines the case considering only $\pi\Delta$ production (i.e. $A$ and $B$ terms in Eq.~\eqref{eq:t_tot}), the blue dash-dotted lines the case considering $\pi\Delta$ and $\pi N^*$ production (i.e. $A$, $B$ and $C$ terms in Eq.~\eqref{eq:t_tot}), the black solid lines the case considering $\pi\Delta$, $\pi N^*$ and $\sigma N$ production (i.e. all the $A$, $B$, $C$ and $D$ terms in Eq.~\eqref{eq:t_tot}), the green dotted lines the case changing the sign of the $g_{\sigma,\pi\pi}$ in Eq. (\ref{eq:gsigmapipi}). The purple dash-dot-dot line in $\frac{{\rm d}\Gamma}{{\rm d}M_{12}}$ distribution corresponds to the case considering only $\sigma N$ production (i.e. $D$ term in Eq.~\eqref{eq:t_tot}) fitted approximately to the $M_{12}$ distribution at low masses.
\label{fig:Fig18}}
\end{figure}

In Fig. \ref{fig:Fig18}, we plot the $\frac{{\rm d}\Gamma}{{\rm d}M_{12}}$, $\frac{{\rm d}\Gamma}{{\rm d}M_{13}}$ and $\frac{{\rm d}\Gamma}{{\rm d}M_{23}}$ mass distributions for the $N^{*+}(1875) \to \pi^+\pi^- p$ decay with 1,2,3 denoting $\pi^+, \pi^-$ and $p$.
Let us first look at the mass distributions considering only $\pi\Delta$ production (red dashed lines in Fig. \ref{fig:Fig18}). We see that for $M_{13}$($\pi^+p$) there is a large signal of the $\Delta^{++}(1232)$ coming from term $B$. The $\Delta(1232)$ is also seen in the $M_{23}$  mass distribution ($\pi^-p$) coming from term $A$. Removing a small background below the $\Delta^0$ peak in the $M_{23}$ distribution, we can see that the strength for $\Delta^{++}$ in the $M_{13}$ distribution is about nine times the one of the $\Delta^0$, as it corresponds to the coefficients in the terms $A$ and $B$ squaring them. The rest of the strength in the $M_{23}$ distribution peaks around $M_{23} = 1500$ MeV, as a consequence of phase space and the weight of the term $B$ non resonant in this channel. The $M_{12}(\pi^+\pi^-)$ mass distribution does not show any resonance and follows phase space weighted by the term $A$ and $B$.

Next we consider the $\pi N^*$ term including in addition the $C$ term in Eq. ~\eqref{eq:t_tot}. The results are shown in Fig.~\ref{fig:Fig18} as the blue dash-dotted lines. We do not see much change except an enhancement of the peak in the $M_{23}$ distribution ($\pi^-p$) corresponding to the $N^*$ excitation by the $C$ term. However, the change is not large. Yet, here we see a possible reason why the the $\pi N^*$ channel has not been claimed experimentally. Indeed, the $\pi\Delta$ mechanism alone already creates a peak in the $M_{23}$ distribution in region of 1500 MeV, which cannot be associated to $\pi N^*$ production. Any $\pi N^*$ production can be easily be attributed to the $\Delta$ production in the $M_{13}(\pi^+p)$ channel.
This has also a consequence in terms of a message: To determine the $\pi N^*$ production one should better look at $\pi\eta N$ production.

We show the results including all the production terms, $A+B+C+D$, in Fig.~\ref{fig:Fig18} as the black solid lines. This includes $\sigma p$ production in addition to the former channels. The results are interesting. Apart from the basic features that we have observed in the former cases, now the $M_{12}(\pi^+\pi^-)$ mass distribution contains a large bump in the region of low invariant $\pi\pi$ masses corresponding to the $\sigma$ production. A smooth extrapolation of the low energy $M_{12}$ distribution with a wide $\sigma$ shape would tell us that about $\frac12$ of the width could be attributed to $\sigma N$ production. To quantify this we have used the $D$ term of Eq. (\ref{def-ABCD}) alone, and roughly adjusted its strength to the low mass region of the $M_{12}$ distribution. This is telling us that an analysis of the mass distributions, due to interference of terms, would provide an apparently larger strength for the $\sigma p $ channel than one would induce from the coupling of the resonance to the different channels, as done in Eqs. (\ref{eq:widths}). Actually, since we are only considering $\frac59$ of the $\pi\Delta$ production in these figures, taking into account the results of Eq. (\ref{eq:widths}), we would be extracting a width of around 7 MeV from this analysis, which would turn into $\frac32 \times 7 \sim 11 \;{\rm MeV}$ if one considers the $\sigma \to \pi^0 \pi^0$ decay also. This is bigger than the 2.3 MeV that we obtained in Eq. (\ref{eq:widths}), and would correspond to a branching fraction of about 15\%.

There is another issue worth considering. In the determination of the couplings there is always a global sign which is arbitrary. The result of the couplings in a coupled channel problem have the relative phase well determined with respect to this one. But the $\pi\pi$ channel is not one coupled to $\pi N$ or $\pi \Delta$. We would like to see what happens if we change the sign of the $g_{\sigma,\pi\pi}$ in Eq. (\ref{eq:gsigmapipi}). The results are shown in Fig. \ref{fig:Fig18} as the green dotted lines and we see that the effects are moderate. One should note that it is precisely in observable that involve interference of the terms that the signs of couplings relative to other signs of, in principle, unrelated couplings can be determined.

\section{Conclusions}
In this work we have complemented the developments of Ref. \cite{sarkar} in which a $3/2^-$ resonance appears around 1875 MeV from the interaction of the $\Sigma^* K$ and $\Delta \pi$ channels. In a first step we introduced the $N^*(1535) \pi$ channel which is produced via
a triangle singularity in which $\Sigma^* K$ is produced, then the $\Sigma^*$ decays to $\Lambda \pi$ and finally the $\Lambda K $ merge to produce the $N^*(1535)$. The interesting observation is that the singularity appears at the same energy of the resonance and then it shows at the same peak and helps stabilize the resonance in the sense that even with a weaker $\Sigma^* K$ interaction the singularity always appears at the same energy.  The other decays channel that we introduced is the $N \sigma$ channel. We also used a triangle diagram to take it into account, taking the $\Delta \pi$ intermediate state, letting the $\Delta$ decay to $N \pi$ and then merging the two pions into a $\sigma$ meson. Then we retake the scheme of \cite {sarkar} adding the two new channels to the original $\Sigma^* K$ and $\Delta \pi$ ones, and with the four coupled channels we study again the resonance, the couplings to the different channels and its decay into these channels. We observe that the partial decay widths of the resonance to
$N^*(1535) \pi$ and $N \sigma$ are not large but measurable. In particular, we observed that the $N \sigma$ channel was much smaller than what is determined experimentally from some experiments. Yet, since channel separation is done from mass distributions, we showed that due to interference with other terms, the $\pi \pi$ mass distribution showed an important enhancement at low invariant masses from where one could extract an appreciable larger fraction of $N \sigma$ than one gets from the couplings, yet smaller than the experimental claims.

   An important part of the work was the study of the $\Sigma^* K$. This channel is not easy to separate in an analysis because the resonance has its mass at the threshold of the channel. In fact no experiment has made claims about this channel. However, we see that the channel is very important in the building of the resonance and that taking into account the width of the resonance and the width of the $\Sigma^*$ we obtained a branching fraction of about 45\%. It is clear that if this channel is omitted in the analysis,  its strength  can easily be attributed to another channel. So, in view of the unavoidable large strength of this channel, we suggest that modern multichannel analyses implementing unitarity in coupled channels are used to revise this resonance. There is a clear example in a related case, where the multichannel analysis provides also a sizeable contribution of a threshold channel, the $N \rho$ in the case of the $N^*(1700) (3/2^-)$ \cite{manley}, where other analyses \cite{elsa} neglect it.

   The determination of the  $N^*(1535) \pi$ channel is also relevant since it will evidence the role of a triangle singularity peaking at the resonance position. Yet, the discussion of the mass distributions in the $\pi^+ \pi^- p$ final state showed that the mass distribution for $N^*(1535) \pi^+, ~N^*(1535) \to \pi^- p $ had the same signature in the $\pi^- p $ mass distribution than that coming from the $\Delta^{++} \pi^-$ excitation mechanisms, where the  $\pi^- p $ is not forming the $\Delta$. This is why if one wishes to determine this channel, the ideal final state should be $\pi \eta N$ not $\pi \pi N$.

  The thorough work conducted here on the building up of the resonance, its decay channels and the mass distributions in the  $\pi \pi N$ channel, together with the discussion above, clearly indicate that a reanalysis of this resonance to the light of the present findings should be most welcome.

\section*{Acknowledgments}
DS acknowledges support from Rajamangala University of Technology Isan, Suranaree University of Technology (SUT) and the Office of the Higher Education Commission under NRU project of Thailand (SUT-COE: High Energy Physics \& Astrophysics) and Thailand Research Fund (TRF) under contract No. MRG5980255.
This work is partly supported by the National Natural Science Foundation of China under Grants
No.~11565007, No.~11647309 and No.~11547307 (WHL).
This work is also partly supported by the Spanish Ministerio
de Economia y Competitividad and European FEDER funds
under the contract number FIS2011-28853-C02-01, FIS2011-
28853-C02-02, FIS2014-57026-REDT, FIS2014-51948-C2-
1-P, and FIS2014-51948-C2-2-P, and the Generalitat Valenciana
in the program Prometeo II-2014/068 (EO). DS is also graceful to IFIC, Universidad de Valencia for hospitality where part of this work was done.

\bibliographystyle{plain}

\end{document}